\documentclass[reqno]{amsproc}
\usepackage{xcolor,amsmath,amssymb,graphicx}
\usepackage[english]{babel}
\usepackage[T1]{fontenc}
\usepackage{amsmath, amscd,amsthm, amsfonts, amssymb}
\usepackage{epsfig}
\usepackage{hhline, latexsym}
\usepackage{float}

\newlength{\dinwidth}
\newlength{\dinmargin}
\newtheorem{definition}{Definition}
\newtheorem{theorem}{Theorem}
\newtheorem{proposition}{Proposition}
\newtheorem{corollary}{Corollary}
\newtheorem{remark}{Remark}
\newtheorem{lemma}{Lemma}

\def\Hgk{{{\mathcal H}_g({\bf k}_{n+m})}}
\def\Hgkd{{\mathcal H}_g^{(2)}({\bf k}_{n+m})}
\def\Hgkp{\tilde{{\mathcal H}}_g^{(2)}({\bf k}_{n+m})}
\def\Hurgk{{\rm Hur}_g({\bf k}_{n+m})}

\def\RS{{\mathcal L}}

\def\be{\begin{equation}}
\def\ee{\end{equation}}
\def\ben{\begin{displaymath}}
\def\een{\end{displaymath}}
\def\baa{\begin{eqnarray}}
\def\eaa{\end{eqnarray}}

\def\ba{\begin{array}}
\def\ea{\end{array}}

\def\ct{\tilde{c}}
\def\lt{\tilde{l}}

\makeatletter
\@addtoreset{equation}{section}
\makeatother

\def\e{\epsilon}

\def\C{{\mathbb C}}
\def\CP1{{\bf CP}^1}

\def\C{{\bf C}}
\def\Z{{\bf Z}}
\def\R{{\bf R}}

\def\Hur{{\rm Hur}}
\def\bHur{\overline{{\rm Hur}}}

\def\a{\alpha}
\def\g{\gamma}

\def\b{\beta}

\def\l{\lambda}

\def\th{\vartheta}
\def\Th{\Theta}

\def\O{\Omega}

\def\Hcal{{\mathcal H}}

\def\f{\frac}
\def\la{\label}

\def\Acal{{\mathcal A}}
\def\p{\partial}

\def\Wcal{{\mathcal W}}

\def\Ccal{{\mathcal C}}
\def\Hmr{{\mathcal H}^{{\bf r}}_{g,m}}
\def\Hmrt{\tilde{\mathcal H}^{{\bf r}}_{g,m}}

\begin{document}
\title{Baker-Akhiezer spinor kernel and  tau-functions on moduli spaces of meromorphic  differentials}

\author{C.~Kalla}  
\address{Universit\'e d'Orl\'eans, UFR Sciences 
MAPMO-UMR
6628, D\'epartement de Math\'ematiques - Route de Chartres
B.P. 6759 - 45067 Orl\'eans cedex 2 
FRANCE  }
\email{caroline.kalla@univ-orleans.fr}

\author{D.~Korotkin}
\address{Department of Mathematics and Statistics, Concordia University,
1455 de Maisonneuve West, Montreal, H3G 1M8  Quebec,  Canada}
\email{korotkin@mathstat.concordia.ca}
\thanks{DK was partially supported by NSERC, FQRNT and CURC}
\maketitle

\vskip0.5cm {\bf Abstract.} 
In this paper we study Baker-Akhiezer spinor kernel on moduli spaces of meromorphic differentials on Riemann surfaces.
We introduce the Baker-Akhiezer tau-function  which is related to both Bergman tau-function (which was studied before in the context of Hurwitz spaces and spaces of holomorphic and quadratic differentials) and KP tau-function on such spaces.

In particular, we derive  variational formulas of Rauch-Ahlfors type on moduli spaces of meromorphic differentials
with prescribed singularities:
we use the system of  homological coordinates, consisting of absolute and relative periods of the 
meromorphic differential, and show how to vary the fundamental objects associated to a Riemann surface (the matrix of $b$-periods, 
normalized Abelian differentials, the Bergman bidifferential, the Szeg\"o kernel and the  Baker-Akhiezer spinor kernel) with respect to these coordinates.   The variational formulas encode   dependence 
both on the moduli of the Riemann surface and on the choice of meromorphic differential (variation of the meromorphic differential while keeping the Riemann surface fixed corresponds to flows of KP type).

Analyzing the global properties of the Bergman and Baker-Akhiezer tau-functions 
 we establish 
relationships between various divisor classes on the moduli spaces.


\vskip0.5cm

\tableofcontents

\section{Introduction}

Meromorphic Abelian differentials on Riemann surfaces play the fundamental role in the theory of integrable systems - from 
algebro-geometric construction of solutions of KdV and KP equations and their generalizations \cite{Dubr,Krich}, when the Riemann surface 
remains fixed - to  Whitham deformations \cite{Krich1} and Frobenius manifolds \cite{Dubr1},
when the Riemann surface is deformed.

The goal of this paper is to derive variational formulas of Ahlfors-Rauch type on moduli spaces of 
meromorphic differentials on Riemann surfaces and to 
apply them to study the Baker-Akhiezer spinor kernel and various tau-functions (the Bergman tau-function,
 as well as the Baker-Akhiezer tau-function which we introduce in this paper) on these spaces.

Consider a Riemann surface $\RS$ and an arbitrary meromorphic differential $v$ on $\RS$ with $n$ poles $x_1,\dots,x_n$ and $m$ zeros  $x_{n+1},\dots, x_{n+m}$;
we fix the multiplicities of the poles and zeros such that the divisor of $v$ is given by 
\be
(v)=\sum_{i=1}^{m+n} k_i x_i\;;
\la{divvint}\ee
 $k_1,\dots,k_n$ are negative and 
$k_{n+1},\dots,k_{n+m}$ are positive; $k_1+\dots+k_{n+m}=2g-2$ is the degree of the canonical divisor.

The moduli space of pairs (Riemann surface $\RS$ of genus $g$, meromorphic differential $v$ with $n$ poles,
 and fixed degrees $k_1,\dots,k_{n+m}$ of divisor $(v)$) is denoted by   $\Hcal_g(k_1,\dots,k_{n+m})$ (briefly $\Hgk$). 

In the traditional approach  used in the theory of integrable systems the system of coordinates on the space 
$\Hgk$ consists of moduli of the Riemann surface punctured at poles $x_1,\dots,x_n$ and coefficients of
singular parts of $v$ near the poles; assuming that all $a$-periods of $v$ vanish, these coordinates determine $\RS$ and $v$ uniquely. The problem with  such coordinate system is that the coefficients of singular parts of $v$ near singularities 
depend on the choice of local parameters near poles. In this paper we use a different coordinate system, which
originates, on one hand, in the theory of Hurwitz spaces, when the local coordinates are given by branch points of
a covering of complex plane; on the other hand, such coordinate system is widely used in the study of the 
Teichm\"uller flow on moduli spaces of holomorphic differentials on Riemann surfaces (see \cite{Zorich}).

Here we study the moduli space of meromorphic differentials using the system of homological coordinates.
The system of homological local coordinates on $\Hgk$  is given by integrals of $v$ along basic
$a$ and $b$ cycles  on $\RS$, integrals of $v$ around its poles, and integrals of $v$ between its zeros.
More precisely, consider the first homology group of the Riemann surface $\RS$ {\it punctured at poles $x_1,\dots,x_n$, relative to the set of zeros $x_{n+1},\dots,x_{m+n}$ of $\RS$}; this homology group
\be
H_1(\RS\setminus\{x_i\}_{i=1}^n;\{x_i\}_{i=n+1}^{n+m})
\la{H1intro}\ee
will be briefly denoted by $H_1$. The  homology group dual to $H_1$ is the group 
\be
H_1(\RS\setminus\{x_i\}_{i=n+1}^{n+m};\{x_i\}_{i=1}^{n})
\la{H1starint}\ee
which we shall briefly denote by  $H_1^*$; this is the first homology group of the surface $\RS$  punctured at {\it zeros} of $v$, 
{\it relative to poles} of $v$. 

Denote by $\{s_i\}_{i=1}^{2g+n+m-2}$ a set of generators of $H_1$; the number
of generators $2g+n+m-2$ coincides with dimension of the moduli space  $\Hgk$ and the coordinates
on  $\Hgk$ can be chosen as $z_i=\int_{s_i} v$.
The dual basis in $H_1^*$ is denoted by $\{s_i^*\}_{i=1}^{2g+n+m-2}$.

If the differential $v$ is exact: $v=df$, where $f$ is a meromorphic function, then all residues of $v$ vanish, as well
as all of its $a$ and $b$-periods. The moduli space of such differentials is just a Hurwitz space, and the non-vanishing homological coordinates coincide with the critical values of function $f$ i.e. with branch points of corresponding branch covering.

Denote by $\O$ the matrix of $b$-periods of $\RS$ computed in some homology basis $(a_\a,b_\a)$ on $\RS$ (the choice of this basis is independent  of the basis 
$\{s_i\}$ mentioned before). The dual basis of holomorphic differentials $u_\a$ is normalized via $\oint_{a_\a} u_\b=\delta_{\a\b}$.

The variational formulas, say, for the matrix of $b$-periods with respect to the homological coordinates look as follows:
\be
\f{\p \O_{\a\b}}{\p z_i}=\int_{s_i^*}\f{u_\a u_\b}{v}\;.
\la{varsigint}
\ee
The formula (\ref{varsigint}) strongly resembles the standard Ahlfors-Rauch formula (see for example \cite{Abikoff}) for variation of the matrix of $b$-periods under change of
conformal structure of the Riemann surface defined by an arbitrary Beltrami differential; in (\ref{varsigint}) the role of ``Beltrami differential'' is played by 
the vector field $1/v$ localized on the contour $s_i^*$. In the case of moduli spaces of holomorphic differentials, the formula (\ref{varsigint}) was proved in \cite{JDG} together with  variational formulas for $u_\a$, the prime-form, canonical bimeromorphic differential and other objects.
In the  special case  of spaces of exact differentials $v=df$ (when all homological coordinates obtained by integration over absolute periods vanish) we are dealing with Hurwitz spaces; the homological coordinates coincide with critical values of the meromorphic function $f$ i.e. with branch points of the corresponding branch covering.

Let us use the differential $v$ to introduce on $\RS$ a coordinate $z(x)$ (for $x$ which don't coincide with poles or zeros of $v$).
Namely,  introduce on $\RS$ a system of cuts which are homologous to $a$ and $b$-cycles, and also fix cuts connecting poles of $v$;
in this way we get the ``fundamental domain''  $\tilde{\RS}$ of $\RS$.
We introduce these cuts in such a way that their homology classes coincide  with the corresponding  generators $s_i^*$ in $H_1^*$.
In the domain $\tilde{\RS}$ the integral 
$$z(x)=\int_{x_{n+1}}^x v$$
(with the initial point coinciding with the first zero  $x_{n+1}$ of $v$) is single-valued, and defines a coordinate near any point $x\not\in (v)$. 
Near poles or zeros of $v$ one can also use $z(x)$ to define a system of so-called {\it distinguished} local coordinates;
these coordinates are discussed below.

Fix a line bundle $\chi$ of degree $g-1$ on $\RS$ and denote by $\delta_\chi$  the corresponding divisor class.
The Szeg\"o kernel on the Riemann surface $\RS$ is the Cauchy reproducing kernel 
acting on sections of
$\chi$.  Introduce two vectors  $p,q\in \R^g$ via $\O p + q= \Acal_{x_0}(\delta_\chi)+ K^{x_0}$, 
where $x_0$ is a basepoint, $ K^{x_0}$ is the vector of Riemann constants and $\Acal$ is the Abel map; assume that $h^0(\chi)=0$ i.e.
 $\th_{pq}(0)\neq 0$, where $\th_{pq}$ is the theta-function with characteristics $[p,q]$.
The vectors $p$ and $q$ are related to holonomies of the line bundle $\chi$ along basic cycles:
$p_\a=-\f{1}{2\pi i}{\rm Hol}(\chi,a_\a)$; $q_\a=\f{1}{2\pi i}{\rm Hol}(\chi,b_\a)$.

The  Szeg\"o kernel $S_{\chi}(x,y)$, which will also be denoted by  $S_{pq}(x,y)$, is defined by the formula 
\be
S_{pq}(x,y)=\f{\th_{pq}(\Acal(x)-\Acal(y))}{\th_{pq}(0)E(x,y)}\;,
\la{defSzegoint}
\ee
where $E(x,y)$ is the prime-form;
 $S_{pq}(x,y)$ is a meromorphic section of 
 $\chi$ with respect to $x$, and of the line bundle  $\chi^{-1}\otimes K $ with respect to $y$, where $K$ is the canonical class.

Assuming that the vectors $p$ and $q$ (i.e. moduli of $\chi$) remain independent of homological coordinates, we prove the following variational formulas for the Szeg\"o kernel with respect to moduli $\{z_i\}$ of the space $\Hgk$:
\be
\f{\p}{\p z_i}{S_{pq}(x,y)}\Big|_{z(x),z(y)}=  \f{1}{4}\int_{t\in s_i^*} \f{W[S_{pq}(x,t)S_{pq}(t,y)]}{v(t)}
\la{varSzegoint}
\ee
where $W$ denotes the Wronskian with respect to variable $t$: locally for any two functions $f(t)$ and $g(t)$  it is given by $f'g-fg'$
(in the theory of integrable system it is also called the first Hirota derivative of $f$ and $g$). The Wronskian
$W[S_{pq}(x,t)S_{pq}(t,y)]$ is a  quadratic differential with respect to $t$. The $z$-coordinate of $x$ and $y$ in (\ref{varSzegoint}) 
is kept constant under differentiation with respect to moduli.

For a fixed Riemann surface one can consider dependence of the Szeg\"o kernel on components of vectors $p$ and $q$ (i.e. 
 on holonomies of the line bundle $\chi$). For derivatives of $S_{pq}$ with respect to $p$ and $q$  the following variational formulas hold:
\be
\f{d}{d p_\a} S_{pq}(x,y)=  - \oint_{b_\a}  S_{pq}(x,t)S_{pq}(t,y)\;,
\hskip0.7cm
\f{d}{d q_\a} S_{pq}(x,y)= - \oint_{a_\a}  S_{pq}(x,t)S_{pq}(t,y)\;,
\la{varSpqint}
\ee
which seem to be new.

The variational formulas (\ref{varSzegoint}) and (\ref{varSpqint}) can be applied to study  the Baker-Akhiezer
spinor kernel, which plays a central role in the theory of integrable systems.
Its definition is inspired by the theory of algebro-geometric solutions of integrable hierarchies of KP type (called Krichever's scheme \cite{Krich}). The name Baker-Akhiezer (BA) is inherited from the Baker-Akhiezer function, the object which first appeared historically. The BA kernel was used 
also in differential geometry (theory of helicoids with handles \cite{Bobenko}). Recently the properties of  BA kernel were studied  in \cite{BE} in the context
of  the  topological recursion of Eynard-Orantin.

The Baker-Akhiezer spinor kernel $S^{(v)}(x,y)$ can be constructed from a triple: Riemann surface $\RS$, meromorphic differential $v$ and a line bundle $\chi$
of degree $g-1$ (a ``twisted'' spinor line bundle). The line bundle $\chi$ determines two vectors $p,q\in \R^g$ as above. Then  $S^{(v)}(x,y)$
 is defined by the formula:
\be
S^{(v)}(x,y) = S_{p_v,q_v}(x,y) \exp\left\{\int_x^y v\right\}\;,
\la{defBAkerINT}
\ee
where vectors $p_v$ and $q_v$ are expressed via periods of the differential $v$ and vectors $p,q$:
$$(p_v)_\a=p_\a-\f{1}{2\pi i}\int_{a_\a}v\;;\hskip0.7cm  (q_v)_\a=q_\a+\f{1}{2\pi i}\int_{b_\a}v\;.$$
The BA  kernel $S^{(v)}(x,y)$ is well-defined at any point of the moduli space 
outside of the divisor defined by the equation   $\th[p_v,q_v](0)=0$.

If the differential $v$ does not have residues (i.e. it is an Abelian differential of second kind), then 
with respect to  variable $x$ the  Baker-Akhiezer spinor is a section of 
the line bundle $\chi$ having one pole (at $x=y$) and essential singularities at
poles of $v$; with respect to $y$ it is a section of $\chi^{-1}\otimes K$.
If $v$ is a differential of third kind (i.e. if it has only simple poles) then the kernel $S^{(v)}(x,y)$ 
does not have essential singularities; instead it has regular singularities at poles of $v$, generically
 connected by branch cuts.

If $v$ has residues at some of its higher order poles, there are also branch cuts connecting the poles with non-trivial residues.

As well as the Szeg\"o kernel, the Baker-Akhiezer kernel is invariant under symplectic transformations of the canonical basis of cycles.
The   Baker-Akhiezer kernel can be interpreted as the Cauchy kernel acting on sections $\psi$ of $\chi$ which are singular at poles of $v$, such that 
$\psi(x) e^{-\int^x v}$ is holomorphic near poles of $v$.

The derivatives of the Baker-Akhiezer kernel with respect to homological coordinates (with vectors $p,q$ kept fixed)
contain the contribution from variational formulas for the Szeg\"o kernel, as well as 
contribution from variation of   the characteristic vectors $p_v$ and $q_v$ 
($p_v-p$ and $q_v-q$ coincide with some of homological coordinates up to a sign and factor of $1/2\pi i$).

The variational formulas on the space $\Hgk$ look as follows:
\be
\f{\p}{\p z_i}{S^{(v)}(x,y)}\Big|_{z(x),z(y)}= \f{1}{4} \int_{t\in s_i^*} \f{W[S^{(v)}(x,t)S^{(v)}(t,y)]}{v(t)} -\f{1}{2\pi i}\int_{t\in \mu(s_i^*)} S^{(v)}(x,t)S^{(v)}(t,y)
\la{varBAint}
\ee
where $\mu$ is the  natural projection of the homology group $H_1(\RS\setminus\{x_i\}_{i=n+1}^{n+m};\{x_i\}_{i=1}^{n})$
(\ref{H1starint}) onto the  homology group $H_1(\RS)$ of the unpunctured Riemann surface $\RS$.

The second term in the integrand takes into account the dependence of $p_v$ and $q_v$ on homological coordinates.

The formula (\ref{varBAint}) has an interesting corollary, first proved in \cite{BE} by a different method. 
Namely, assume that the
Riemann surface $\RS$ remains fixed, while the differential $v$ is changed to
$v+\e v_{a}$ where 
$$v_{a}(x)= \f{B(x,a)}{v(a)} $$ 
is a (normalized)  meromorphic differential of the second kind having a pole of 
second order at $a\not\in (v)$. Denote by $\delta_{a}$ the derivative
with respect to $\e$ at $\e=0$. 

Variations corresponding to different points of the Riemann surface satisfy the following algebra:
$$
[\delta_{a},\delta_{b}]=\f{B(a,b)}{v(a)v(b)}(\delta_{a}-\delta_{b})\;,
$$
whose structure constants are given by the canonical bidifferential $B$.

Acting by $\delta_{a}$ on the BA kernel we see that
the first term in the integrand of the   formula (\ref{varBAint}) does not contribute when  $\delta_{a}$ is applied to $S^{(v)}(x,y)$; the second term of the integrand gives the following equation (which in slightly different form first appeared in \cite{BE}):
\be
\delta_{a} S^{(v)}(x,y) = \frac{S^{(v)}(x,a) S^{(v)}(a,y)}{v(a)}\;.
\la{BEcorrected}
\ee

For a fixed Riemann surface the  variational formula (\ref{varBAint}) implies the 
equations for the Baker-Akhiezer kernel, which are equivalent to the integrable hierarchy of KP type, being written in terms of standard ``times''
of these hierarchies. If the Riemann surface also varies, these formulas describe the dependence of the Baker-Akhiezer kernel on moduli of the Riemann surface,
providing a simple and geometrically transparent alternative to an  earlier construction by Grinevich and Orlov
 \cite{GrinOrlov}.

Another  application of the variational formulas on spaces of meromorphic differentials is the theory of Bergman tau-function.  So far the notion of Bergman tau-function was introduced for two kinds of spaces, namely, for Hurwitz spaces, and the spaces of holomorphic differentials over Riemann surfaces. In the context of Hurwitz spaces the Bergman tau-function coincides with the isomonodromic tau-function of matrix Riemann-Hilbert problem associated to Hurwitz Frobenius manifolds \cite{Dubr,Crelle}. The Bergman tau-function is also a non-trivial ingredient of the Jimbo-Miwa tau-function of Riemann-Hilbert problems with quasi-permutation monodromy groups \cite{Annalen}; it appears 
also in large N expansion of partition function of hermitian matrix models \cite{EKK}.

The Bergman tau-function on spaces of holomorphic differentials arises in the problem of holomorphic factorization 
of determinant of Laplace operator on Riemann surfaces with flat metrics with conical singularities and trivial holonomy
\cite{JDG}; from the point of view of conformal field theory this tau-function can be interpreted as chiral scalar partition function of
free bosonic field theory on a Riemann surface (with respect to such flat metric), see for example \cite{VerVer}. 

The  moduli spaces of {\it holomorphic}  differentials over Riemann surfaces have also recently attracted a lot of attention of
experts working in the area of dynamical systems: it turns out that geometry of these spaces is very closely related to fundamental
properties of the Teichm\"uller flow (the sum of  Lyapunov exponents of such flow, see \cite{EKZ,Zorich} and references therein). 
In particular, according to the fundamental theorem of \cite{Zorich}, the sum of Lyapunov exponents can be expressed via the Hodge class and other divisor classes on such spaces.
Study of global properties of the Bergman tau-function allows to get new relations between various divisor classes on moduli spaces of
holomorphic differentials \cite{MRL}; these relations turn out to be useful in deriving the formulas for the sum of Lyapunov exponents of the Teichm\"uller flow \cite{EKZ}.

Geometrically, the Bergman tau-function on spaces of holomorphic differentials, as well as on Hurwitz spaces, is a section of Hodge line bundle multiplied by a degree of the tautological line bundle.
 A detailed study of analytical properties of the tau-function allows to express the Hodge class in terms of other 
geometrically important divisor classes on spaces of admissible covers (the compactifications of Hurwitz spaces) \cite {Annalen} and spaces of holomorphic differentials \cite{MRL}.

In this paper we extend the notion of Bergman tau-function to spaces of meromorphic differentials. 
 In particular, this allows to
extend the results about determinant of Laplace operator on flat Riemann surfaces to the physically important 
case of Mandelstam diagram (such diagrams correspond to spaces of differentials of third kind).
Since spaces of meromorphic differentials contain Hurwitz spaces and spaces of holomorphic differentials as natural subspaces,
this provides a unifying framework for results of \cite{JDG,MRL,Advances}.
Moreover,  the Bergman tau-function on spaces of meromorphic differentials should be more suitable for use in the
theory of random matrices and Chekhov-Eynard-Orantin topological recursion than its version on Hurwitz spaces used
in \cite{EKK}.

In analogy to \cite{Advances,MRL}, introduce on $\RS$ the  meromorphic differential $Q_v$, which is
essentially given by the zero order term of the asymptotics of the  Bergman bidifferential
$B(x,y)=d_x d_y\log E(x,y)$ near diagonal:
\be
Q_v(x)=\frac{1}{v(x)}\left( B(x,y)-\f{v(x)v(y)}{(\int_x^y v)^2}\right)\Big|_{y=x}
\la{defQvint}
\ee
(up to the factor $1/v(x)$, $Q_v$ is given by the Bergman projective connection computed in the local parameter $\zeta(x)=\int^xv$).


The differential $Q_v$ has several important properties. First, $Q_v(x)$ is holomorphic at all points $x$ where $v$ 
is holomorphic and non-vanishing (i.e. for $x\not\in (v)$). At zero  $x_i$ of order $k_i$ ($i=n+1,\dots,n+m$) the differential $Q_v$ has pole of order $k_i+2$. At pole $x_i$ of $v$ of order greater than 2 (i.e. if $-k_i>2$ for some $i=1,\dots,n$) the differential $Q_v$ has zero of order $k_i-2$. At pole of $v$ of  order $2$ the differential $Q_v$ has zero of order at least 1. Finally, if at some point
$v$ has simple pole with residue $r$, then $Q_v$ also has simple pole at this point, and its residue equals 
$-(12r)^{-1}$.

When some poles of $v$ are simple, the formalism of Bergman tau-function becomes more complicated in comparison with the case when all poles of $v$ are of higher order. Therefore, we consider the case of spaces of second kind differentials
(assuming that all poles of $v$ are of order greater or equal than $2$ and all residues of $v$ vanish) separately;
this case turns out to be a relatively straightforward generalization of corresponding formalism for spaces  of holomorphic differentials and Hurwitz spaces.

Suppose that in $(v)=\sum_{i=1}^{n+m} k_i x_i$ all $k_i\neq -1$, i.e. all poles of $v$ have order 2 and higher, and, moreover, all residues at these  poles vanish.
In that case homological coordinates on $\Hgk$ are given by integrals of $v$ over some basis in the homology group of $\RS$, relative to the set of zeros of $v$:
\be
H_1(\RS ;\{x_i\}_{i=n+1}^{n+m})
\la{H1int2}
\ee
A set of generators $\{s_i\}$ in the group (\ref{H1int2}) can be chosen to consist of canonical $a$ and $b$-cycles $(a_\a,b_\a)$ on $\RS$, and contours $l_i$ which connect the 
first zero $x_{n+1}$ of $v$ with zeros $x_{n+i}$. The set of dual generators $s_i^*$ then consists of cycles $-b_\alpha$, $a_\alpha$ and small circles around
$x_{n+2},\dots,x_{n+m}$.

Then  the Bergman tau-function on the given stratum of the space of differentials of the second kind is defined as follows:
\be
d\log\tau_B(\RS,v)= -\f{1}{2\pi i} \sum_{i=1}^{2g+m-1} \left(\int_{s_i^*} Q_v\right) d\left(\int_{s_i} v\right)
\la{deftauint}
\ee
The 1-form in the right-hand side of (\ref{deftauint}) is closed as a corollary of  variational formulas  on the space $\Hgk$; this implies that the tau-function 
can be defined locally in a neighborhood of any point of $\Hgk$. 

As well as in the cases of holomorphic differentials and Hurwitz spaces, the $\tau$-function $\tau_B(\RS,v)$ 
can be computed explicitly in terms of theta-functions and prime-forms. Moreover, for certain power 
$N$ depending on multiplicities $\{k_i\}$, $\tau_B^N(\RS,v)$ turns out to be a section of the product of 
 $N$th power of the Hodge line bundle over  
$\Hgk$ and certain power of the tautological line bundle. Since  $\tau_B$ is non-singular and non-vanishing on the 
(open part of) the moduli space, this allows to express the Hodge class via the tautological class in the Picard group of
(the open part of) the  moduli space. We consider in detail the simplest non-trivial genus one situation, when the differential
has only one pole of certain multiplicity $k$, and one zero of the same multiplicity; in this case a
 degree of the  Bergman tau-function is a modular form of level $k$.

By combining the Bergman tau-function with theta-function and certain exponential term, we introduce the object which we denote by 
$\tau_{BA}$ and  call the Bergman-Akhiezer tau-function:

\be
\tau_{BA}(\RS,v,\chi)=  e^{-\pi i \langle p_v, q_v\rangle} \tau_B^{-1/2} \th[p_v,q_v](0)
\la{tauBAint}
\ee
The Baker-Akhiezer tau-function  (taken to an appropriate power to get rid of certain roots of unity) is a modular-invariand object: it is independent of the choice of canonical basis of cycles. While the Bergman tau-function is defined from regularized (near diagonal $x=y$)  Bergman canonical bidifferential $B(x,y)$, 
the Baker-Akhiezer tau-function can
be defined intrinsically via regularization of the  Baker-Akhiezer spinor kernel at the  diagonal. In the case of exact differentials $v$ the Baker-Akhiezer tau-function coincides with
the isomonodromic tau-function of Riemann-Hilbert problems with quasi-permutation monodromy matrices \cite{Birk,Annalen}.

The Bergman tau-function on  moduli spaces of differentials of third kind is of particular interest due to its 
connection with chiral partition function of free bosons on Mandelstam diagrams \cite{DH}. 
However, in that case the natural definition of the Bergman tau-function becomes
less invariant than in the cases of differentials of first and second kind: it can be defined for fixed values 
of the residues of the differentials, and the definition depends in particular on the choice of the ``first'' zero of the differential.
Otherwise the formalism resembles a lot the simpler cases of differentials of first and second kind. Here we treat in detail the 
first non-trivial case of genus 0 and 4 simple poles.

The paper is organized as follows. In Section \ref{varformsec} we introduce all necessary objects, define homological coordinates and derive variational formulas for holomorphic differentials, canonical bimeromorphic differential and Szeg\"o kernel 
 on moduli spaces of meromorphic differentials of an arbitrary type.
In Section \ref{BAsection} we introduce the Baker-Akhiezer kernel and study its variations with respect to homological coordinates; we specially consider variations preserving the conformal class of the Riemann surface, since these variations enter the 
KP-type hierarchies  corresponding to the BA kernel. In Section \ref{sectionTAU} we define the Bergman and Baker-Akhiezer tau-functions on spaces of meromorphic differentials of second kind, study their basic properties, discuss the corresponding line bundles and deduce new relations in the Picard group of the moduli spaces of differentials of second kind.
In Section \ref{sectionTHIRD} we define Bergman tau-function on spaces of differentials of third kind, discuss ambiguities arising in this case, and study the simplest non-trivial example in detail.

\section{Variational formulas on spaces of meromorphic differentials in homological coordinates}

\la{varformsec}

\subsection{Auxiliary objects}

Here  we introduce a few auxiliary objects. On  a compact Riemann surface $\RS$ of genus
$g$ introduce a canonical basis of cycles $(a_\a,b_\a)$ in  $H_1(\RS,\Z)$. Denote by $B(x,y)$ for $x,y\in \RS$ the Bergman bidifferential,
which is the symmetric bimeromorphic differential on $\RS$ having quadratic pole with biresidue 1 on the diagonal and vanishing $a$-periods. 
The bidifferential $B$ is expressed via the the prime-form $E(x,y)$ as follows: $B(x,y)=d_x d_y \log E(x,y)$. Consider a basis of holomorphic differentials
$u_\a$ on $\RS$ normalized as follows:
\be
\oint_{a_\a}u_{\beta}=\delta_{\a\b}
\ee
The $b$-periods of $u_\a$ give the matrix of $b$-periods $\O$ of $\RS$:
\be
\O_{\a\b}=\oint_{b_\a} u_\b
\ee
Choosing some local coordinate $\xi$ near the diagonal $\{x=y\}\subset \RS\times \RS$,  we have the following expansion of $B(x,y)$ near the 
diagonal:
\be
B(x,y)=\left(\f{1}{(\xi(x)-\xi(y))^2}+\f{S_B(\xi(x))}{6}
+O((\xi(x)-\xi(y))^2)\right)d\xi(x) d\xi(y),
\la{asW}
\ee
where $S_B$ is a projective connection on $\RS$ called the {\em Bergman projective connection}.

If two canonical bases of cycles on $\RS$, $\{a_\a',b_\a'\}_{\a=1}^g$ and 
$\{a_\a,b_\a\}_{\a=1}^g$
 are related by
a matrix 
\be
\sigma=
\left(\begin{array}{cc} d & c\\
b & a \end{array}\right)\in Sp(2g,\Z)\;,
\la{symtrans}
\ee
then the corresponding Bergman bidifferentials are related as follows (see item 4 on page 21 of \cite{Fay73}):
\be
B^{\sigma}(x,y)=B(x,y)-2\pi i\sum_{\a,\b=1}^g  [(c\O+d)^{-1} c]_{\a\b} u_\a(x) u_\b(y)\;.
\la{BsB}
\ee
  
The matrix of $b$-periods $\Omega^\sigma$ corresponding to the new canonical basis of cycles, is related to $\Omega$ as follows:
\be
\Omega^\sigma=(a\O+b) (c\O+d)^{-1}
\la{sympBper}
\ee

Let us introduce the theta-function $\th_{pq}(z,\Omega)$ (see \cite{Tata1} for its definition and properties):
\be
\th_{pq}(z,\Omega) = e^{\pi i \langle p, \Omega p\rangle +2\pi i \langle p , z+q\rangle}\th(z+\Omega p+q)
\la{defthchar}
\ee
where the standard theta-function $\th(z)$ is given by
\be
\th(z)=\sum_{m\in \Z^g} {\rm exp}(\pi i \langle \Omega m, m\rangle +2\pi i \langle \Omega z, m\rangle)
\la{defthz}
\ee

 We shall also use the following modified theta-function with vanishing argument $z$:
\be
\tilde{\th}_{pq}(0) = e^{-\pi i \langle p, q\rangle} \th_{pq}(0,\Omega)\;;
\la{defthtil}
\ee
in \cite{Tata1} (formula (5.3)) this function is denoted by $\th^\alpha[p,q](0)$; the advantage of $\tilde{\th}_{pq}$ in comparison with  $\th_{pq}(0,\Omega)$ is that it's transformation law under change of symplectic homology basis is  simpler than the transformation of  $\th_{pq}(0)$.

The next key object we introduce in this section is the Szeg\"o kernel.

Let $\chi$ be an arbitrary line bundle of degree $g-1$; denote corresponding divisor class by $\delta_\chi$
 and define two vectors $p,q\in \R^g$ via relation 
\be
\O p + q= \Acal_{x_0}(\delta_\chi)+ K^{x_0}
\la{pqchi}
\ee
where $K^{x_0}$ is the vector of Riemann constants corresponding to an arbitrary basepoint $x_0\in\RS$;  $\Acal_{x_0}$ is the Abel map corresponding to the same base point.

Components of vectors $p$ and $q$ are related to holonomies of the line bundle $\chi$ along the basic cycles:
\be
p_\a= -\f{1}{2\pi i}\log {\rm Hol}(\chi, a_\a)\;,\hskip0.7cm
q_\a= \f{1}{2\pi i}\log {\rm Hol}(\chi, b_\a)
\la{defholon}
\ee

Assume that $\chi$ does not have holomorphic sections (i.e. $h^0(\chi)=0$); then Riemann's theorem implies  $\th_{pq}(0)\neq 0$ and we can define the  Szeg\"o kernel $S_{\chi}(x,y)$ corresponding to the twisted spin line bundle $\chi$ via the following formula 
\footnote{In the sequel we shall use both notations  $S_{\chi}(x,y)$  and  $S_{pq}(x,y)$ for the Szeg\"o kernel. In principle,  
 $S_{\chi}(x,y)$
is a preferable notation: since the line bundle $\chi$ is an object chosen independently of canonical basis of cycles, the Szeg\"o kernel  $S_{\chi}(x,y)$ is manifestly independent of the  choice of canonical basis. On the other hand, vectors $p$ and $q$ do depend on this choice, and notation $S_{pq}(x,y)$ hides this invariance.}
\be
S_\chi(x,y):=S_{pq}(x,y)=\f{\th_{pq}(\Acal(x)-\Acal(y))}{\th_{pq}(0)E(x,y)}
\la{defSzego}
\ee

 The Szeg\"o kernel is the Cauchy reproducing kernel in the line bundle 
$\chi$; $S_{pq}(x,y)$ is a section of the
line bundle $\chi$ with respect to $x$, and of the line bundle  $\chi^{-1}\otimes K$ ($K$ is the canonical line bundle) with respect to $y$.

From transformation of the vector of Riemann constants under the change of canonical basis of cycles given by (\ref{symtrans}) (see for example lemma 1.5 of 
\cite{Fay92}) and (\ref{pqchi}), 
 it follows that the
 vectors $p$ and $q$  transform as follows:
\be
\left[\ba{c} p^\sigma \\ q^\sigma \ea\right]=\left(\ba{cc} d & -c \\ -b & a \ea\right) \left(\ba{c} p \\ q \ea\right)+
\f{1}{2}\left[\ba{c} (c^t d)_0   \\ (a^t b)_0   \ea\right]\;.
\la{transchar}
\ee
where the index ``0'' denotes the diagonal part of the corresponding matrix.

The ``modified'' theta-function $\tilde{\th}_{pq}(0)$ transforms under the symplectic change (\ref{symtrans}) of canonical basis as follows ((5.4) in \cite{Tata1}):
\be
\tilde{\th}_{p^\sigma q^\sigma}(0|\Omega^\sigma)=\gamma(\sigma) {\rm det}^{1/2}(c\Omega+d) \tilde{\th}_{pq}(0|\Omega)
\la{transth}
\ee  
where $\gamma(\sigma)$ is an 8th root of unity: $\gamma^8=1$.

Let us now introduce the holomorphic differential
\be
 w(x)=\sum_{\a=1}^g \p_{\a} \log\th_{pq}(0) u_\a(x)
\la{defw}\ee

In terms of differential $w$ we can write the following simple formulas for derivatives of the modified theta-function $\tilde{\th}_{pq}$ with respect to components of characteristic vectors $p$ and $q$:

\begin{lemma} \la{varmodth}
The following formulas hold:
\be
\f{d}{d p_\a} \log \tilde{\th}_{pq}(0) = \pi i q_\a + \oint_{b_\a} w
\la{modthp}
\ee
\be
\f{d}{d q_\a} \log \tilde{\th}_{pq}(0) = -\pi i p_\a + \oint_{a_\a} w
\la{modthq}
\ee
\end{lemma}
The {\it proof} of this lemma immediately follows from the definitions (\ref{defthz}), (\ref{defthchar}) and (\ref{defthtil}) of theta-function with characteristics and the modified theta-function 
$\tilde{\th}_{pq}$.

$\Box$

The next  proposition (strangely enough we did not find it in existing literature) describes variation of  Szeg\"o kernel with respect to vectors $p$ and $q$ (i.e. it describes variation of $S_{\chi}(x,y)$ with respect to holonomies of the line bundle $\chi$).

\begin{proposition}\la{thvarpq}
The following variational formulas for the Szeg\"o kernel with respect to components of 
characteristic vectors $p$ and $q$ hold:
\be
\f{d}{d p_\a} S_{pq}(x,y)=  - \oint_{b_\a}  S_{pq}(x,t)S_{pq}(t,y)
\la{varSp}
\ee
\be
\f{d}{d q_\a} S_{pq}(x,y)= - \oint_{a_\a}  S_{pq}(x,t)S_{pq}(t,y)
\la{varSq}
\ee
\end{proposition}
{\it Proof.}  Let us consider (\ref{varSp}). The Szeg\"o kernel has the following transformation law under analytical continuation along cycle $a_\a$:
\be
S_{pq}(x+a_\a,y)= e^{2 \pi i p_\a} S_{pq}(x,y)
\la{translaw}
\ee
Differentiating (\ref{translaw})  with respect to $p_\a$, we get:
$$
\f{d}{d p_\a} S_{pq}(x+a_\a,y)= 2 \pi i e^{2 \pi i p_\a}  S_{pq}(x,y) + e^{2 \pi i p_\a} \f{d}{d p_\a} S_{pq}(x,y)
$$
which, due to (\ref{translaw}), can also be written as

\be
\f{d}{d p_\a} S_{pq}(x_+,y)= 2 \pi i   S_{pq}(x_+,y) + e^{2 \pi i p_\a} \f{d}{d p_\a} S_{pq}(x_-,y)
\la{translawdif}
\ee
where $x_-=x$ and $x_+=x+a_\a$ are points on different sides of the cycle $b_\a$. 
Relation (\ref{translawdif}) shows that $\f{d}{d p_\a} S_{pq}(x,y)$ is a section of $\chi $  (with respect to  $x$) which has an additive jump
 $2 \pi i  S_{pq}(x,y)$ on the cycle $b_\a$. Since $ S_{pq}(x,y)$ is  itself nothing but the Cauchy kernel in the bundle  $\chi $,
it can be used to get an alternative expression for  $\f{d}{d p_\a} S_{pq}(x,y)$. Namely, integrating the additive jump function $2 \pi i e^{2 \pi i p_\a}  S_{pq}(x,t)$ multiplied with  the Cauchy kernel $ S_{pq}(t,y)$ along the cycle $b_\a$, we see that  $\f{d}{d p_\a} S_{pq}(x,y)$ is given by (\ref{varSp}).

Similar analysis for  $\f{d}{d q_\a} S_{pq}(x,y)$ taking into account the automorphy property $S_{pq}(x+b_\a,y)= e^{-2 \pi i q_\a} S_{pq}(x,y)$ with respect to
$b$-cycle, gives (\ref{varSq}) (the sign in (\ref{varSq}) is the same as in (\ref{varSp}): one minus sign arises from  ``-'' in the automorphy factor, and another
minus sign arises from skew-symmetry of the intersection index $a_\a\circ b_\a$).

$\Box$

\begin{remark}\rm 
Notice that the variational formulas (\ref{varSp}),  (\ref{varSq}) are consistent with the asymptotic of the Szeg\"o 
kernel near diagonal, as $y \to x$ (\cite{Fay92}, p.29), which in a local coordinate $\xi$ can be written as follows:
\be
S_{pq}(x,y)=\left\{\f{1}{\xi(y)-\xi(x)}+ C_0(x) + C_1(x) (\xi(y)-\xi(x))\right\}\sqrt{d\xi(x)}\sqrt{d\xi(y)}
\la{asdia}
\ee
where 
\be
C_0(x)= \f{w(x)}{d\xi(x)}\;,
\la{a0}\ee
(the differential $w(x)$ is given by (\ref{defw}));
\be
C_1(x) = \f{1}{12} S_B(\xi(x)) +\f{1}{2} \sum_{\a=1}^g \p_\a\log\th_{pq}(0) \f{d}{d\xi(x)}\left(\f{u_\a(x)}{d\xi(x)}\right)+
\f{1}{2}\sum_{\a,\b=1}^g \p_\a\p_\b\log\th_{pq}(0) \f{u_\a(x) u_\b(x)}{(d\xi(x))^2}\;;
\la{a1}\ee
$\p_{\a}$ denotes derivative of theta-function with respect to its $\a$th argument and $\xi$ is an arbitrary local coordinate near $x$.
Let us for simplicity look at  (\ref{varSq}); the characteristic $q$ corresponds to the shift of the argument of theta-function by $q$ and,
therefore, $\p_{q_\a}\th_{pq}(0)=\p_{\a}\th_{pq}(0)$.
Differentiating asymptotics (\ref{asdia}) with respect to $q_\b$, we get that, as $y\to x$,
\be
\p_{q_\b}S_{pq}(x,y)\big|_{y=x}= - \sum_{\a=1}^g \p_\a\p_\b\log\th_{pq}(0) u_\a(x)\;.
\la{derasim}
\ee
On the other hand, taking the limit $y\to x$ in  the formula (\ref{varSq}), 
using the well-known relation (formula (39) of \cite{Fay73})
\be
S_{pq}(x,t)S_{pq}(t,x)=- B(x,t)- \sum_{\a,\b=1}^g \p_\a\p_\b\log\th_{pq}(0)\,
 u_\a(x) u_\b(t)
\la{SSB}\ee
where $B(x,t)$ is the Bergman bidifferential, and taking into account that $B$ has vanishing $a$-periods with respect to both variables, we come to the same formula
(\ref{derasim}).

Similarly one can verify that the asymptotics (\ref{asdia})  is consistent with 
(\ref{varSp}) as $y\to x$; the calculation in this case is slightly more complicated since $b$-periods of $B(x,y)$ don't vanish; up to factor of $2\pi i$ they are given by the
corresponding holomorphic differentials $u_\a$.
\end{remark}

Let us now consider some Abelian differential $v$ on $\RS$; it could be a differential of either first, second or third kind on $\RS$;
it can be also a meromorphic differential of mixed type (i.e having both poles of higher order and residues).

Introduce the meromorphic differential $Q_v$, which is
 given by the zero order term of the asymptotics of the  Bergman bidifferential on the diagonal: 
$$
Q_v=\frac{  B_{reg}(x,x)}{v(x)}\;,
$$
where the regularization is done using the differential $v$, i.e., 
\be
Q_v(x)=\frac{1}{v(x)}\left( B(x,y)-\f{v(x)v(y)}{(\int_{x}^y v)^2}\right)\Big|_{y=x}\;.
\la{defQv}
\ee

To give an alternative definition of  $Q_v$  we  denote by $\{\cdot,\cdot\}$ the usual Schwarzian derivative.
Define the following meromorphic projective connection associated with differential $v$:
\be
S_v:= \left\{ \int^x v, \xi(x)\right\}\equiv
\frac{v''}{v}-\frac{3}{2}\left(\frac{v'}{v}\right)^2\;,
\la{defSv}
\ee
where prime denotes the derivative with respect to a local coordinate $\xi$. 

Since for a given 1-differential $v$,  $S_v$ is a meromorphic projective connection on $\RS$, 
the difference $S_B-S_v$ is a meromorphic quadratic differential. 
 Dividing $S_B-S_v$ by $v$ we get the differential $Q_v$:
\be
Q_v:= \f{1}{6}\f{S_B-S_v}{v}
\la{defQv1}
\ee
The meromorphic Abelian differential $Q_v$,  constructed from the  (holomorphic or meromorphic) Abelian differential $v$ 
 plays the key role in the construction of the Bergman tau-function. Notice that $Q_v$ is determined by $v$ and a choice of canonical basis of cycles on $\RS$.

\begin{remark}\rm
Using the definition of Schwarzian derivative (\ref{defSv})  it is easy to verify that at pole  $x_i$ ($i=1,\dots,n$) of order $-k_i$  the differential $Q_v$ has 
zero of order $k_i-2$ for poles of order $3$ and higher (i.e. when $-k_i\geq 3$). At a pole of $v$ of second order, i.e. when $k_i=-2$, the differential $Q_v$ has zero of order at least 1.
Finally, at a simple pole of $v$ with residue $r$, the differential $Q_v$ also has a simple pole, and its residue equals $-1/(12r)$. 
At the zero $x_{n+i}$ of order $k_{n+i}$, the differential  $Q_v$ has pole of order $k_{n+i}+2$. 
\end{remark}

As a corollary of (\ref{BsB}), $Q_v$ transforms as follows under symplectic transformation $\sigma$:
\be
Q_v^{\sigma}(x)=Q_v(x) - 12\pi i\sum_{\a,\b=1}^g  [(c\O+d)^{-1} c]_{\a\b} \f{u_\a(x) u_\b(x)}{v(x)}
\la{transQv}
\ee

Another object we are going to use  in the sequel is   given by the following expression:
\be
\Ccal(x)=\frac{1}{\Wcal(x)}\sum_{\a_1, \dots,
\a_g=1}^g
\frac{\partial^g\th(K^x)}{\p z_{\a_1}\dots \p z_{\a_g}}
u_{\a_1}\dots u_{\a_g}(x)\;,
\la{Ccal}\ee
where
\be
\Wcal(x):= {\rm \det}_{1\leq \a, \b\leq g}||u_\b^{(\a-1)}(x)||
\la{Wronks}
\ee
is the Wronskian determinant of the basic holomorphic differentials, and $K^x$ is the vector of Riemann constants with initial point $x$.
The expression (\ref{Ccal}) is a $n(1-n)/2$-differential which is non-single-valued on $\RS$; $\Ccal(x)$ does not have any zeros or poles:
the zeros of the Wronskian in the denominator are the Weierstrass points; the numerator has at the Weierstrass points zeros of the same order.

In the case of  genus $1$ the $x$-dependence in (\ref{Ccal}) drops out and $\Ccal(x)$ turns into $\th'((\O+1)/2)$.

\subsection{Variational formulas on spaces of meromorphic differentials}
\la{Vfmd}

Following the notations introduced in the introduction, we consider the space $\Hgk$ of meromorphic differentials $v$ on Riemann surface $\RS$ of genus $g$ such that
the divisor of $v$ is given by $(v)=\sum_{i=1}^{m+n} k_i x_i$, with $k_1,\dots,k_n<0$ and $k_{n+1},\dots,k_{n+m}>0$.

Considering the homology group of $\RS$ punctured at poles of $v$, relative to the set of zeros of $v$, which is denoted by 
\be
H_1(\RS\setminus\{x_i\}_{i=1}^n;\{x_i\}_{i=n+1}^{n+m})\;,
\la{relhol1}
\ee
we can choose a set of generators $\{s_i\}_{i=1}^{2g+m+n-2}$ as follows:
$$
s_\a= a_\a\;,\;\;  s_{\a+g}= b_\a \hskip0.5cm \a=1,\dots,g\;; \hskip0.5cm
s_{2g+k}=c_{k+1}\;, k=1,\dots,n-1\;,
$$
\be
s_{2g+n-1+k}= l_{n+1+k}\;,\hskip0.5cm k=1,\dots,m-1\;,
\la{defsmerom}
\ee
where 
$c_2,\dots,c_{n}$ are small contours around poles $x_2,\dots,x_{n}$;
$l_{n+2},\dots,l_{n+m}$ are contours connecting the ``first'' zero $x_{n+1}$ with other zeros $x_{n+2},\dots,x_{n+m}$.
 
 The homological coordinates on $\Hgk$ are defined as integrals of $v$ over $\{s_i\}$:
\be
z_i=\int_{s_i} v\;,\hskip0.5cm i=1,\dots, 2g+n+m-2
\la{homcoormain}
\ee

The homology group dual to (\ref{relhol1}) is the homology group of $\RS$, punctured at {\it zeros} of $v$, relative to the set of {\it poles} of $v$; it is denoted by 
\be
H_1(\RS\setminus\{x_i\}_{i=n+1}^{n+m};\{x_i\}_{i=1}^{n})\;.
\la{defsstarmer}
\ee
The dual to (\ref{defsmerom}) set  of generators $\{s_i^*\}_{i=1}^{2g+n+m-2}$ in the group (\ref{defsstarmer}) is given by 
 $$
s^*_\a= -b_\a\;,\;\;  s^*_{\a+g}= a_\a \hskip0.5cm \a=1,\dots,g\;; \hskip0.5cm
s^*_{2g+k}= -\lt_{k+1}\;,\hskip0.5cm k=1,\dots,n-1\;
$$
\be
s^*_{2g+n-1+k}= \ct_{n+1+k}\;,\hskip0.5cm k=1,\dots,m-1
\la{dualcont2}
\ee
where $\lt_2,\dots,\lt_{n}$ are contours connecting the ``first'' pole $x_1$ with other poles $x_2,\dots,x_n$, respectively; $\ct_{n+2},\dots,\ct_{m+n}$ are small
circles around the zeros $x_{n+2},\dots, x_{n+m}$.

The variational formulas for the matrix of $b$-periods, basic holomorphic differentials, Bergman bidifferential and bidifferential 
$Q_v$
formally look  analogous to the case of the space of
holomorphic differentials (see Theorem 3 of \cite{JDG}). 

To write down these formulas we introduce  on $\RS$ a system of cuts homologous to $a$- and $b$-cycles to get the fundamental polygon $\hat{\RS}$; inside
of  $\hat{\RS}$ we also introduce branch cuts connecting poles of $v$ with non-vanishing residues; these branch cuts are assumed to
start at $x_1$, i.e. they connect $x_1$ with $x_2,\dots,x_n$; denote them by the same letters $\tilde{l}_2,\dots,\tilde{l}_n$ 
as their homology classes in (\ref{dualcont2}). In this way we get a domain $\widehat{\RS}_0$ where 
the Abelian integral $z(x)=\int_{x_{n+1}}^x v$ is single-valued.

Now we are in a position to formulate the following theorem, which gives variational formulas on $\Hgk$ with respect to homological coordinates 
$z_i=\int_{s_i}v$.

\begin{theorem}\la{thvarfo}

The variational formulas for matrix of $b$-periods $\O$, normalized holomorphic differentials $u_\a$, Bergman bidifferential $B(x,y)$ and differential $Q_v$
 on the moduli  space $\Hgk$ look as follows (in the cases of $u_\a$, $B(x,y)$ and $Q_v(x)$ the 
coordinates $z(x)$ and $z(y)$ remain fixed under differentiation):
\be
\f{d \O_{\a\b}}{d z_i} =\int_{s_i^*}\f{u_\a u_\b}{v}
\la{varO2}
\ee 
\be
\f{d u_{\a}(x)}{d z_i}\Big|_{z(x)=const} =\f{1}{2\pi i}\int_{t\in s_i^*}\f{u_\a(t)B(x,t)}{v(t)}
\la{varva2}
\ee 
\be
\f{d B(x,y)}{d z_i}\Big|_{z(x),z(y)=const} 
=\f{1}{2\pi i}\int_{t\in s_i^*}\f{B(x,t)B(y,t)}{v(t)}
\la{varBxy2}
\ee 
\be
\f{d Q_v(x)}{d z_i}\Big|_{z(x)=const} 
=\f{1}{2\pi i}\int_{t\in s_i^*}\f{B^2(x,t)}{v(t)}
\la{varQv2}
\ee 
\end{theorem}

The {\it proof} of this theorem follows the proof of Theorem 3 of \cite{JDG}. The proof of  variational formulas with respect to coordinates
$\int_{a_\a}v$,  $\int_{b_\a}v$ and $\int_{x_{n+1}}^{x_{n+j}}v$ repeat the corresponding proof from \cite{JDG} verbatim.

Formally the new ingredient of the variational formulas on moduli spaces of {\it meromorphic} differentials is
 the variational formulas with respect to residues $r_2,\dots, r_n$ of
$v$ at $x_2,\dots,x_n$. The logic of the proof of these variational formulas is also parallel to  \cite{JDG}. Namely, the residue at $x_1$ equals 
$-(r_2+\dots+r_n)$. Consider, for example, the formula (\ref{varva2}) with $z_i=r_2$. The differential in the left-hand-side of this formula has 
an additive jump on the contour $\tilde{l}_2$, which is given by  the differential $d(u_\a/v)$. Since the Bergman bidifferential plays the role of Cauchy kernel (with second order pole on the diagonal), such differential with additive jump on  $\tilde{l}_2$  can be written as integral
of $\f{1}{2\pi i}\f{u_\a(t)B(x,t)}{v(t)}$ over $\tilde{l}_2$, which gives the right-hand side of (\ref{varva2}).

$\Box$

The variational formulas for Szeg\"o kernel on moduli spaces of holomorphic differentials were not considered in \cite{JDG}, 
but in the context of Hurwitz spaces the variation of $S_{pq}(x,y)$ with respect to a critical value of meromorphic function was given 
in Theorem 3  of \cite{Annalen}.

Generalizing this theorem from Hurwitz spaces to an arbitrary stratum $\Hgk$ of the space of meromorphic differentials we get the following 
theorem. In this theorem  $W_t$ denotes the Wronskian with respect to variable $t$: locally for any two functions $f(t)$ and $g(t)$  it is given by $f'g-fg'$
(in the theory of integrable system it is also called the first Hirota derivative of $f$ and $g$).

\begin{theorem}\la{derSzke}
Derivatives of Szeg\"o kernel with respect to homological coordinates $z_i$ on $\Hgk$ are given by
\be
\f{\p}{\p z_i}{S_{pq}(x,y)}\Big|_{z(x),z(y)}= \f{1}{4}\int_{t\in s_i^*} \f{W_t[S_{pq}(x,t),\;S_{pq}(t,y)]}{v(t)}
\la{varSzego}
\ee
\end{theorem}

We notice that the Wronskian $W_t[S_{pq}(x,t),\, S_{pq}(t,y)]$ is a quadratic differential with respect to $t$; dividing it by $v(t)$ we get a meromorphic differential on $\RS$ (with respect to $t$). 

The {\it proof} of (\ref{varSzego}), when coordinate $z_i$ is  one of integrals between zeros of $v$: $ z_i=\int_{x_{n+1}}^{x_{n+i}} v$, coincides with the proof of Theorem 3 of \cite{Annalen}, where variational formula for Szeg\"o kernel with respect to branch points of a branched covering were proved.

The variational formulas  with respect to other coordinates are obtained by an elementary combination of the proof of Th.3 of \cite{JDG} and Th.3 of \cite{Annalen}.

\begin{remark}\rm (Variational formulas in fatgraph picture).
Variational formulas given by Theorems  \ref{thvarfo} and \ref{derSzke} admit a nice reformulation in terms of fatgraphs with chosen orientation of edges.
Namely, assume  that $n\geq 1$ i.e.  $v$ has at least one pole. Then there exists a (non-unique) fatgraph $\Gamma$ with $m$ vertices at zeros
of $v$ such that the  union of faces of  $\Gamma$ coincides with $C$, and inside of each face one has exactly one pole of $v$ (i.e. there are $n$ faces). 
The number of edges then equals $2g-2+n+m$ and coincides with the dimension of the homology group (\ref{relhol1}). Thus one can choose the (homology classes of oriented)
 edges as a basis $\{s_i\}$ in (\ref{relhol1}). Let us now consider the dual graph $\tilde{\Gamma}$; it has $n$ vertices which coincide with poles of $v$. Each face of $\tilde{\Gamma}$
contains exactly one zero of $v$. The number of edges of $\Gamma$ coincides with the number of edges of 
$\tilde{\Gamma}$;
to each (oriented) edge $s_i$ of $\Gamma$ one can uniquely assign its dual, which is 
 an oriented edge $s_i^*$ of
$\tilde{\Gamma}$. The set of oriented edges $\{s_i^*\}$ is a basis in the dual homology group (\ref{defsstarmer}).  By construction of the dual graph the intersection index of $s_i^*$ and $s_j$ equals $\delta_{ij}$. Therefore, one can use the edges of the fatgraph $\Gamma$  to define the 
homological coordinates $z_i=\int_{s_i} v$; in the right hand sides of all variational formulas above one then has to  integrate over the corresponding dual edge $s_i^*$ of the dual fatgraph $\tilde{\Gamma}$.
\end{remark}

\subsection{Deformations preserving the complex structure of the base Riemann surface.}

Generically, when $m> g-1$, the dimension of $\Hgk$ is greater than $3g-3+n$ (which is the dimension of the moduli space 
of Riemann surfaces of genus $g$ with $n$ marked points (for $g>1$)).
The natural question in this case is how to find those combinations of the vector fields $\p_{z_i}$ on the moduli space $\Hgk$  which do not change the 
complex structure of the Riemann surface $C$ and positions of $n$ marked points (the poles of $v$). The simplest answer can be given in  the case when all zeros of the differential $v$ are simple,
i.e. $k_{n+1}=\dots=k_{n+m}=1$. Then $m=2g-2-\sum_{i=1}^n k_i$ and the dimension of $\Hgk$ equals $4g-4+n-\sum_{i=1}^n k_i$.
Since the dimension of ${\mathcal M}_{g,n}$ equals $3g-3+n$, there exist $g-1-\sum_{i=1}^n k_i$ linearly independent combinations of
$\p_{z_i}$ which preserve the complex structure of the Riemann surface $C$ and marked points $x_1,\dots, x_n$.
 
The number $g-1-\sum_{i=1}^n k_i$  equals the dimension of the linear vector space of Abelian differentials $w$ 
such that $(w)\geq \sum_{i=1}^n k_i x_i$
($w$ is an arbitrary linear combination of holomorphic differentials and normalized differentials of 
second  and third kind with poles at poles of $v$ of order less or equal to $-k_i$). 

\begin{proposition}
Let $w$ be an arbitrary Abelian differential on $C$ such that $(w)\geq \sum_{i=1}^n k_i x_i$. Then the vector field 
\be
V_w := \sum_{i=1}^{2g+n+m-2}\left(\int_{s_i}w\right) \f{\p}{\p z_i}
\la{pw}
\ee
preserves the moduli of the Riemann surface $C$ and positions of points $x_1,\dots,x_n$.
\end{proposition}

{\it Proof.} Since we assumed that all zeros of $v$ are simple, all zeros of the differential 
$v^\e =v+\epsilon w$ are also simple for 
$\e$ small enough. The differential $v^\e$ is defined on the same Riemann surface and has the same set of poles as $v$.
Therefore, the vector field $\f{d}{d\e}$ does not change the complex structure of $C$ and positions of poles. On the other hand, expressing this vector field via
$\f{\p}{\p z_i}$, we see that $\f{d}{d\e}=V_w$.

$\Box$

One can alternatively check that the matrix of $b$-periods does not change along the orbits of the vector fields (\ref{pw}); this can be done using variational formulas (\ref{varO2}) and Riemann bilinear identity.

The coefficients of vector fields (\ref{pw}) which are given by integrals of $w$ over all generators $s_i$, can be easily computed for any of basic differentials $w$ (i.e. holomorphic normalized differentials and normalized differentials of second and third kind with poles at $x_1,\dots,x_n$).

A more non-trivial consideration is required to treat the case of zeros of $v$ of an arbitrary multiplicity,
since after adding of a small perturbation to $v$ the zeros of higher order generically split into several 
simple zeros, and the perturbed differential leaves the original stratum of the moduli space.

Below we consider the case $n=0$, i.e. the case of spaces of holomorphic differentials with zeros 
at $x_1,\dots, x_m$ of multiplicities $k_1,\dots,k_m$ such that $k_1+\dots+k_m=2g-2$.

According to Torelli's theorem, to determine which vector fields do not change the conformal structure of the Riemann surface $C$, one needs to find those linear combinations $V_j$ of the 
vector  fields $\p_{z_i}$,
which annihilate the matrix of $b$-periods  $\Omega$.
The number of such combinations is $m-g+2$ (this number arises as difference of the dimension of the stratum of the moduli space of holomorphic differentials ($2g+m-1$) and the dimension of the moduli space of Riemann surfaces ($3g-3$)  i.e. it equals $g$ when all zeros are simple and $m=2g-2$);
if $m\leq g-2$, then such vector fields are generically absent.

Let us write $V_j$ as a linear combinations of $\p_{z_i}$:

\be
V_j=\sum_{\g=1}^g \left\{A_j^\g \f{\p}{\p (\int_{a_\g} v)} + B_j^\g \f{\p}{\p (\int_{b_\g} v)}\right\}
 + \sum_{k=2}^{m} C_j^k \f{\p}{\p (\int_{l_k} v)}
\la{Vj}
\ee
where the matrices $A$, $B$ and $C$ will be determined from conditions $V_j(\Omega)=0$.

According to variational formulas (\ref{varO2}), 
\be
V_j(\O_{\a\b})=\sum_{\g=1}^g \left\{-A_j^\g \int_{b_\g}\f{v_\a v_\b}{v} + B_j^\g\int_{a_\g}\f{v_\a v_\b}{v}
\right\} + 2\pi i \sum_{k=2}^{m} C_j^k \;{\rm res}|_{x_k}\f{v_\a v_\b}{v}
\la{VjO}
\ee

Let us now assume that the curve $C$ is non-hyperelliptic (according to \cite{KonZor}, hyperelliptic curves span a separate connected component of moduli spaces of holomorphic differentials). Then linear combinations of products 
$v_\a v_\b$ of holomorphic normalized differentials span the whole space of holomorphic quadratic differentials 
on $C$. In turn, if $q$ is a holomorphic quadratic differential on $C$, then $u=q/v$ is a meromorphic Abelian differential with
poles of order $k_i$ at $x_i$; moreover, by choosing an appropriate $q$  one can get an arbitrary meromorphic differential $u$ such that $(u)\geq -\sum_{i=1}^m k_i x_i$. A basis in the space of such differentials $u$ consists of basic normalized differentials $u_1,\dots, u_g$, normalized differentials of the
third kind $w_{x_1 x_i}$ for $i=2,\dots,m$ and normalized differentials of the second kind $v_{x_i}^{(s)}$
for $i=1,\dots,m$, $s=2,\dots,k_i$, corresponding to some choice of local parameters near $x_i$ (notice that this basis consists of $g+(m-1) +\sum_{i=1}^{m} (k_i-1)=3g-3$
differentials, which coincides with the dimension of the space of holomorphic quadratic differentials $q$).
(We recall that the differentials of second kind are defined as follows: 
$v_{x_i}^{(s)}$ has pole of order $s$ at $x_i$ with the singular part given by $\xi_i^{-s} d\xi_i$, where 
$\xi_i$ is a local parameter near $x_i$, and all $a$-periods of $v_{x_i}^{(s)}$ vanish).

Now, requiring that
\be
\sum_{\g=1}^g \left\{-A_j^\g \int_{b_\g} u + B_j^\g\int_{a_\g} u
\right\} + 2\pi i \sum_{k=2}^{m} C_j^k \;{\rm res}|_{x_k} u =0
\la{peru}
\ee
for any $u$ from this list, we get a set of relations for coefficients of matrices $A$, $B$ and $C$.
The periods of differentials $u_\a$, $w_{x_1 x_i}$ and $v_{x_i}^{(s)}$ look as follows:
\be
\oint_{a_\b} u_\a =\delta_{\a\b}\;,\hskip0.7cm \oint_{b_\b} u_\a =\O_{\a\b}\;,\hskip0.7cm
{\rm res}|_{x_k} \{w_{x_1 x_k} \}=1
\la{peru1}
\ee
\be
\oint_{b_\b} w_{x_1 x_i}=2\pi i \int_{l_i} u_\b,\hskip0.7cm
\oint_{b_\b} v_{x_i}^{(s)} = {2\pi i}\f{u_\b^{(s-2)}(x_i)}{(s-1)!}\;,
\la{peru2}
\ee
where $u_\b^{(s-2)}(x_i)$ stands for $\f{d^{s-2}}{d\xi_i^{s-2}}\left(\f{u_\b}{d\xi_i}\right)\Big|_{\xi_i=0}$.
All other periods of these differentials which enter (\ref{peru}), vanish.

Let us introduce the column vector of holomorphic normalized differentials ${\bf u}=(u_1,\dots,u_g)^t$.

Substituting (\ref{peru1}) and (\ref{peru2}) into (\ref{peru}), we come to the following 
\begin{proposition}\la{degenhol}
Suppose that the point $(C,v)$ does not belong to the hyperelliptic component of the corresponding 
stratum of the moduli space of holomorphic differentials.
Let $A$ be an arbitrary matrix of size $(m-g+2)\times g$ and rank $m-g+2$ containing all b-periods of differentials
$v_{x_i}^{(s)}$ in its kernel, i.e.
\be
{\bf u}^{(s)} (x_i)\in {\rm ker}\,A\;\hskip0.7cm i=1,\dots,m,\;\;\; s=0,\dots k_i-2
\la{kernel}
\ee
(the number of these equations equals $2g-2-m$). 

Then vector fields (\ref{Vj}), where matrices $B$ and $C$ are given by
\be
B= A\Omega\;\hskip0.7cm
C= A \left(\int_{l_2} {\bf u}\;,\dots,\int_{l_m} {\bf u}\right)\;,
\ee
 preserve the 
complex structure of the Riemann surface $C$. 
\end{proposition}

Notice that conditions (\ref{kernel}) are empty if all zeros of $v$ are simple; in that case $A$ can be chosen to be the $g\times g$ unit matrix. This situation was already covered by the previous proposition.

\section{Baker-Akhiezer spinor kernel on moduli spaces of meromorphic differentials}
\la{BAsection}

\subsection{Definition and basic properties}
\la{defvarfo}

Consider a line bundle $\chi$ of degree $g-1$ over $\RS$ and define corresponding characteristic  vectors  $p, q\in \R^g$ via  (\ref{pqchi}).

Define now two new vectors $p_v, q_v\in \C^g$, which coincide with vectors $p$ and $q$ up to periods of the meromorphic differential $v$: 
\be
(p_v)_\a=p_\a - \f{1}{2\pi i}\int_{a_\a} v\;;\hskip0.7cm 
(q_v)_\a=q_\a + \f{1}{2\pi i}\int_{b_\a} v
\la{pvqv}\ee
Introduce the line bundle $\chi_v$ of degree $g-1$, depending on the differential $v$, such that the corresponding divisor class $\delta_{\chi_v}$ satisfies
$\Acal_{x_0}(\delta_{\chi_v})+K_{x_0}=\Omega p_v +q_v$.

The Baker-Akhiezer spinor kernel is defined by the triple (Riemann surface, meromorphic differential $v$,  line bundle $\chi$ of degree $g-1$); it
 is expressed via the Szeg\"o kernel  by the formula:
\be
S^{(v)}(x,y) = S_{p_v,q_v}(x,y) \exp\left\{\int_y^x v\right\}\;,
\la{defBAker}
\ee
or, equivalently,
\be
S^{(v)}(x,y) = \f{\th[p_v,q_v](\Acal(x)-\Acal(y))}{\th[p_v,q_v](0) E(x,y)} \exp\left\{\int_y^x v\right\}\;.
\la{defBAker1}
\ee
As well as the Szeg\"o kernel, the Baker-Akhiezer (briefly BA) kernel is invariant under  symplectic transformations of the canonical basis of cycles.
We notice that vectors $p_v$ and $q_v$, given by (\ref{pvqv}), generically have complex-valued components, 
since we don't impose here any reality conditions on periods of the differential $v$.

The BA kernel is well-defined if $h^0(\chi_v)=0$ i.e. the theta-function in the denominator is non-vanishing:
\be
\th[p_v,q_v](0)\neq 0\;;
\la{nonsing}
\ee
therefore, the BA kernel is defined on the moduli space $\Hgk$ outside of the divisor defined  by equation $\th[p_v,q_v](0)=0$, or, equivalently,
\be
\O p_v+q_v\in (\Th)\;,
\la{thetadiv}
\ee
where $(\Th)$ is the theta-divisor on Jacobian $J(\RS)$.

 The divisor in the moduli space defined by relation 
(\ref{thetadiv}) for fixed vectors $p$ and $q$ (i.e. for fixed moduli  of the line bundle $\chi $),  will be denoted by $(\Theta^{\chi})$:
\be
(\RS,v)\in (\Theta^{\chi}) \Leftrightarrow \O p_v+q_v\in (\Th)\;.
\la{defThetabig}
\ee

If the differential $v$ does not have residues (i.e. it is an Abelian differential of second kind), then 
the Baker-Akhiezer  kernel (\ref{defBAker}) is a section of the line bundle  $\chi$ with respect to variable $x$, and  
$\chi^{-1}\otimes K$ with respect to variable $y$. With respect to each variable  $x$ and $y$ the kernel $S^{(v)}(x,y)$
has essential singularities at poles  of $v$. Moreover,  $S^{(v)}(x,y)$ has simple pole on the diagonal $x=y$.

If $v$ is a differential of third kind having only simple poles then the kernel $S^{(v)}(x,y)$ 
does not have essential singularities; instead it has regular singularities at poles of $v$, connected by branch cuts.

If $v$ has residues at some of its poles, there are also branch cuts connecting these poles with non-trivial residues.

For a fixed Riemann surface $\RS$, the variational formulas (\ref{varSp}) and (\ref{varSq}) which describe dependence of the Szeg\"o kernel on
characteristic vectors $p$ and $q$, have the following interesting  corollary. Namely, assume that the
Riemann surface $\RS$ and vectors $p,q$ remain unchanged, while the differential $v$ is changed to
$v_\e:=v+\e v_{a}$ where 
\be
v_{a}(x)= \f{B(x,a)}{v(a)} 
\la{defvx0}
\ee
is the meromorphic differential of the second kind having a pole of 
second order at point $a$ (not coinciding with any of poles or zeros of $v$), and vanishing periods over cycles $a_\a$.

Denote by $\delta_{a}$ the derivative 
with respect to $\e$ at $\e=0$. Obviously,
\be
\delta_{a} v(x)= \f{B(x,a)}{v(a)}\;.
\la{deltav}\ee

The operators $\delta_{a}$ are vector fields on space of differentials of second kind for a fixed Riemann surface $\RS$
(clearly, $\delta_{a}$ don't generically preserve a given stratum of the moduli space  since addition of $\f{B(x,a)}{v(a)}$
changes the pole structure of $v$).

\begin{proposition}
Vector fields (\ref{deltav}) satisfy the following commutation relations:
\be
[\delta_{a},\delta_{b}]=\f{B(a,b)}{v(a)v(b)}(\delta_{a}-\delta_{b})\;.
\la{commrel}
\ee
\end{proposition}
{\it Proof.}  To verify relation (\ref{commrel}) on any functional on the space of meromorphic differentials it's sufficient to act 
by (\ref{commrel}) on $v$ itself. 
We have
$\delta_{a} v(x)= \f{B(x,a)}{v(a)}$ and
$$
\delta_{b}\delta_{a} v(x)=-\f{B(a,b)}{v(a)v(b)}\f{B(x,a)}{v(a)}=-\f{B(a,b)}{v(a)v(b)}\delta_{a} v(x)\;,
$$
which implies (\ref{commrel}).

A simple computation shows that the algebra defined by (\ref{commrel}) satisfies the Jacobi identity.

$\Box$

The following corollary of Proposition \ref{thvarpq} shows that such variation of the Baker-Akhiezer kernel is given by a 
simple formula (called in \cite{BE} the ``self-replication property''):

\begin{proposition}\la{recipro1}
Assume that $a$ does not coincide   with singularities of $v$. Then 
the $\delta_{a}$ variation of the Baker-Akhiezer kernel is given by the formula
\be
\delta_{a} S^{(v)}(x,y) = \frac{S^{(v)}(x,a) S^{(v)}(a,y)}{v(a)}\;.
\la{BE1}
\ee
\end{proposition}
{\it Proof.}  
Denote $v_\e:= v+\e v_{a}$. For the BA kernel 
$$S^{(v_\e)}(x,y)= S_{p_{v_\e},q_{v_\e}}(x,y) \exp\left\{\int_y^x v_\e\right\} $$
we have
\be
\f{d}{d\e}(p_{v_\e})_\a=- \f{1}{2\pi i} \int_{a_\a} {v_{a}}\;,\hskip0.7cm
\f{d}{d\e}(q_{v_\e})_\a= \f{1}{2\pi i} \int_{b_\a} {v_{a}}\;.
\ee
Therefore, taking into account the formulas (\ref{varSp}), (\ref{varSq}) for derivatives of Szeg\"o kernel with respect to 
components of characteristic vectors, we get:
$$
\f{d}{d\e}\Big|_{\e=0} S^{(v_\e)}(x,y)= e^{\int_y^x v}\left(\int_y^x v_{a} \right)  S_{p_v,q_v}(x,y) 
$$
\be
+\f{1}{2\pi i}\sum_{\a=1}^g \left[\left\{ -\left( \oint_{a_\a} v_{a} \right) \left(\oint_{b_\a} S^{(v)}(x,t) S^{(v)}(t,y) \right)
+ \left( \oint_{b_\a} v_{a} \right) \left(\oint_{a_\a}  S^{(v)}(x,t) S^{(v)}(t,y)\right)\right\}\right]\;.
\la{vares}\ee
The sum over $\alpha$ in this formula can be rewritten via sum of residues of the differential 
$$-\left(\int^t v_{a}\right) S^{(v)}(x,t) S^{(v)}(t,y)$$ inside of the fundamental polygon of $\RS$ (with respect to variable $t$). This differential has poles with residues at 
three points: $t= a$, $t=x$ and $t=y$. Sum of the residues at $t=x$ and $t=y$ cancels against the first term in (\ref{vares}), while the
residue at $t=a$ gives the right-hand side of (\ref{BE1}).

$\Box$

The next proposition describes the ``Schlesinger transformation'' of the BA kernel: we assume that the differential $v$ is changed to $v+ w_{ab}$, where 
$w_{ab}$ is a normalized (all $a$-periods vanish) meromorphic differential of third kind $w_{ab}$
with residues $+1$ and $-1$ at $a$ and $b$, respectively:
 \begin{proposition} \la{Schlesinger}
The following identity holds:
\be
S^{(v+w_{ab})}(x,y)=S^{(v)}(x,y)-\f{S^{(v)}(x,b) S^{(v)}(a,y)}{S^{(v)}(a,b)}\;.
\la{schles1}
\ee
\end{proposition}

{\it Proof.} The formula (\ref{schles1}) can be easily derived from the standard Fay's determinantal identity (formula (43) of \cite{Fay73}) for Szeg\"o kernel, which for the case of $2\times 2$ matrices (Fay's trisecant identity)
looks as follows:
$$
S_{pq}(x,y) S_{pq}(a,b)-S_{pq}(x,b) S_{pq}(a,y)
$$
\be
=\f{\th[p,q](\Acal(x+a-y-b))}{\th[p,q](0)}\f{E(x,a)E(y,b)}{E(x,y)E(a,b)E(x,b)E(y,a)}\;.
\la{fayid}
\ee

Using the definition of the BA kernel, the relation (\ref{schles1})  can be immediately deduced from (\ref{fayid})
$\Box$

Iteration of (\ref{schles1}) gives determinantal formula for $S^{(v+w_{a_1b_1}+\dots+w_{a_n b_n})}$ in terms of $S^{(v)}$; the same formula can be derived from 
$(n+1)\times(n+1)$ Fay's determinant  identity for Szeg\"o kernel.

The ``Schlesinger transformation'' formula (\ref{schles1})
 implies another interesting property  of the Baker-Akhiezer kernel - the   ``reciprocity'':
\begin{proposition} \la{recipro2}
The Baker-Akhiezer kernel satisfies the following ``reciprocity'' property under addition to $v$ of a normalized (all $a$-periods vanish) meromorphic differential of third kind $w_{ab}$
with residues $+1$ and $-1$ at $a$ and $b$, respectively:
\be
\f{S^{(v+w_{ab})}(x,y)}{S^{(v)}(x,y)}=\f{ S^{(v+w_{xy})}(a,b)}{ S^{(v)}(a,b)}
\la{recip}
\ee
where $x,y,a,b$ are four arbitrary points on $\RS$.
\end{proposition}

The property (\ref{recip}) can be also easily checked directly, without using  (\ref{schles1}).

\subsection{Variational formulas for the Baker-Akhiezer kernel}

Propositions \ref{recipro1} and \ref{recipro2} describe dependence of the Baker-Akhiezer kernel on the choice of meromorphic differential $v$ 
when the Riemann surface $\RS$ remains fixed. Using variation formula (\ref{varSzego}) of the Szeg\"o kernel with respect to full moduli on the space $\Hgk$ we can get the complete set of  variational formulas for the Baker-Akhiezer kernel $S_v$ on this space. Namely, combining the
definition (\ref{defBAker}), (\ref{pvqv}) of the BA kernel with  variational formulas (\ref{varSzego}) of Szeg\"o kernel with respect to 
moduli coordinates $z_i$ and components of characteristic vectors $p_\a$, $q_\a$, we get  Theorem \ref{BAvar} below.

Let us first introduce the natural  projection $\mu$ from the relative homology group $H_1(\RS\setminus\{x_i\}_{i=n+1}^{n+m};\{x_i\}_{i=1}^{n})$
 of $\RS$ punctured at zeros of $v$, relative to the set of poles of $v$, to the absolute homology group $H_1(\RS,\Z)$ of $\RS$ (in terms of the generators (\ref{dualcont2})
this projection looks as follows: it maps closed cycles $a_\a$, $b_\a$  to corresponding generators of the absolute homology group, while all cycles  $\tilde{l}_k$ and $\tilde{c}_{n+1+k}$ 
are mapped to $0$).
The projection $\mu$ enters the next formula since variation of Szeg\"o kernel with respect to characteristic vectors (\ref{varSp}),  
(\ref{varSq})
 involves only integrals along cycles from the absolute homology group $H_1(\RS,\Z)$.

Then the formulas  (\ref{varSp}),  
(\ref{varSq}), together with the variational formula (\ref{varSzego}), immediately imply the following theorem:

\begin{theorem}\la{BAvar}
The  variational formulas  for the Baker-Akhiezer kernel on the space $\Hgk$ look as follows:
\be
\f{\p}{\p z_i}{S^{(v)}(x,y)}\Big|_{z(x),z(y)}=\f{1}{4} \int_{t\in s_i^*} \f{W_t[S^{(v)}(x,t),\;S^{(v)}(t,y)]}{v(t)} - \f{1}{2\pi i}\int_{t\in \mu(s_i^*)} S^{(v)}(x,t)S^{(v)}(t,y)
\la{varBAmain}
\ee
where, as usual, the differentiation with respect to homological coordinates $z_i=\int_{s_i}v$ in the left-hand side is performed, assuming that 
the integrals $z(x)=\int_{x_{n+1}}^x v$ and  $z(y)=\int_{x_{n+1}}^y v$ remain constant; $W_t$ stands for Wronskian with respect to variable $t$.
\end{theorem}

The second part of the variational formula (\ref{varBAmain}) for BA kernel on spaces of meromorphic differentials encodes dependence of  BA kernel on the 
differential $v$ for fixed Riemann surface $\RS$.

The first term in  (\ref{varBAmain}) encodes dependence of  BA kernel on   moduli  of the Riemann surface $\RS$. 

In fact, the first term in the right-hand side of (\ref{varBAmain}) gives a very simple alternative to 
 construction of \cite{GrinOrlov} where  the
dependence of the Baker-Akhiezer function on moduli of the underlying Riemann surface was first addressed.


Theorem \ref{BAvar} describes dependence of the BA kernel on coordinates on moduli space of meromorphic differentials for fixed vectors $p,q\in\R^g$. For a fixed point of the moduli space, dependence of the BA kernel  on components of the vectors $p$ and $q$ follows from formulas (\ref{varSp}),  (\ref{varSq})   for derivatives of the Szeg\"o kernel with respect to components of the characteristic vectors.
It is given by the following proposition:
\begin{proposition}\la{varpqSv}
The following variational formulas for the Baker-Akhiezer kernel with respect to components of 
real characteristic vectors $p$ and $q$ defining the line bundle $\chi$ hold:
\be
\f{d}{d p_\a} S^{(v)}(x,y)=  - \oint_{b_\a}  S^{(v)}(x,t)S^{(v)}(t,y)\;,
\la{varSpBA}
\ee
\be
\f{d}{d q_\a} S^{(v)}(x,y)= - \oint_{a_\a}  S^{(v)}(x,t)S^{(v)}(t,y)\;.
\la{varSqBA}
\ee
\end{proposition}

\subsection{Baker-Akhiezer kernel and KP times}

Let us consider the case when the differential $v$ has only one pole (of degree $k:=-k_1$) 
at a point $a\in \RS$. 
This is an object familiar from the theory of 
algebro-geometric solutions of Kadomtsev-Petviashvili (KP) equation (called the Krichever scheme). 

Let us choose in a neighborhood of $a$ 
some local coordinate $\xi(x)$ and write the singular part of $v$ near $a$ in the local parameter $\xi$:
\be
v(\xi)= - \left(\f{\l_1}{\xi^2}+\f{2\l_2}{\xi^3}+\dots+\f{(k-1)\l_{k-1} }{\xi^k}+ O(1)\right)d\xi\;.
\la{vx0}
\ee
Then the differential $v$ 
can be represented as follows:
\be
v=\sum_{j=1}^{k-1} \l_j v_j\;,
\la{vvj}
\ee
where the differential of second kind $v_j$ has the following singular part at $a$:
\be
v_j(x)=\left(-\frac{j}{\xi^{j+1}(x)} + O(1)\right)d\xi(x)\;.
\la{vjdef}
\ee
In contrast to the standard Krichever's scheme we don't normalize differential $v$ by condition of vanishing of its $a$-periods.
We notice that the decomposition (\ref{vvj}) of the differential $v$ depends on the choice of the local parameter $\xi$ near $a$.

The variables $\l_j$ are the so-called ``KP-times''; and it's instructive to see how variational formulas for the BA kernel 
$S^{(v)}$ with respect to the KP times arise  within our formalism. Namely, the application of the chain rule to formulas (\ref{varSp}), (\ref{varSq})
gives the following corollary:

\begin{corollary}
Derivatives of the BA kernel with respect to KP times are given by:
\be
\f{d}{d \l_n} S^{(v)}(x,y) =  {\rm res}\Big|_{t=a}\left\{\frac{S^{(v)}(x,t)S^{(v)}(t,y)}{\xi^n(t)}\right\}\;, \;\;\; n=1,\dots, k-1
\la{derBAKP}
\ee
\end{corollary}
The {\it Proof} is a simple exercise on the use of Riemann bilinear identity, analogous to proof of Proposition \ref{recipro1}.
Namely, analogously to (\ref{vares}), we can write, as a corollary of (\ref{varSp}), (\ref{varSq}):
\be
\f{d}{d \l_n} S^{(v)}(x,y) = S^{(v)}(x,y) \left(\int_y^x v_n\right)+
\sum_{t\in \{x,y,a\}} {\rm res}\left\{\left(\int^t v_n\right) S^{(v)}(x,t) S^{(v)}(t,y)\right\}\;;
\la{dsds}
\ee
sum of the residues at $t=x$ and $t=y$ cancels against the first term in (\ref{dsds}); the residue at $t=a$ gives the 
right-hand side 
of (\ref{derBAKP}).

$\Box$

\section{Tau-functions on spaces of differentials of second kind}
\la{sectionTAU}

Here we are going to introduce the Bergman tau-function on moduli spaces of meromorphic differentials, and study its relationship to KP tau-function and Baker-Akhiezer
kernel.

In defining and studying the  Bergman tau-function we encounter additional difficulties if the differential $v$
has non-vanishing residues. These difficulties are avoided by considering spaces of differentials of second kind.  On the other hand, 
spaces of differentials of second kind can also be viewed as 
natural generalizations of Hurwitz spaces (i.e. spaces of meromorphic functions of given degree
on Riemann surfaces of given genus): namely, the Hurwitz spaces can be considered as spaces of {\it exact} meromorphic differentials of second kind.
In other words, the Hurwitz spaces are subspaces of  spaces of meromorphic differentials of second kind characterized by condition of
vanishing of all absolute periods. On the other hand, when the  poles are absent, the spaces of meromorphic differentials turn into
spaces of holomorphic differentials. 
Therefore, by studying the tau-function on spaces of meromorphic differentials of second kind, we provide a unifying framework for tau-functions 
on Hurwitz spaces (which were originally introduced and studied in \cite{Annalen,IMRN,Advances}) and tau-functions on spaces of holomorphic differentials
(which were introduced and studied in \cite{JDG,MRL}).

Therefore, we don't consider here the important case when differential $v$ is of third kind i.e. has only simple poles (this case is
of particular interest in string theory since the corresponding Bergman tau-function in
this case  is the chiral partition function of free bosons on Riemann surfaces).
The definition and study of the Bergman tau-function in this case encounters  additional difficulties,
which we shall discuss in the next section.

Let us denote by $\Hgkd$ the subspace of $\Hgk$ corresponding to all vanishing residues at  the poles
$\{x_i\}_{i=1}^n$.

The dimension of the space $\Hgkd$ equals $2g+m-1$ (in particular, if all zeros are simple, their number equals
$m=2g-2+k_1+\dots+k_n$, and the dimension equals $4g-3+\sum_{i=1}^n k_i$, coinciding with the dimension of the moduli space of holomorphic differentials with
simple zeros if $n=0$.

The set of homological coordinates on  $\Hgkd$ is constructed in the same way as the set of 
homological coordinates
on spaces of general differentials, excluding the integrals around poles of $v$: we introduce a set of generators in the  
homology group  $H_1(\RS;\{x_i\}_{i=n+1}^{m+n})$  
of the Riemann surface $\RS$ relative to the set of zeros of $v$:
\be
s_\a= a_\a\;,\;\;  s_{\a+g}= b_\a \hskip0.5cm \a=1,\dots,g\;; \hskip0.5cm
s_{2g+k-n-1}= l_k\;,\hskip0.5cm k=n+2,\dots,n+m\;;
\la{defsmer2}
\ee
where $l_{n+2},\dots,l_{n+m}$ are contours connecting the ``first'' zero $x_{n+1}$ of $v$  with  other zeros.

As well as in the general case of the spaces $\Hgk$, the homological coordinates are defined as integrals of $v$ over $\{s_i\}$:
$z_i=\int_{s_i} v $.

The Hurwitz spaces (i.e. spaces of meromorphic functions of given degree and given multiplicities of poles and critical points) are subspaces of
spaces $\Hgkd$ which correspond to exact differentials $v$: $v=df$, where $f$ is the meromorphic function.
Thus the Hurwitz spaces can be described in terms of homological coordinates on  $\Hgkd$ as subspaces
defined by vanishing of all absolute periods of the differential $v$:
\be
\int_{a_\a}v=\int_{b_\a}v=0\;.
\la{condHur}
\ee

The dual set  of generators in the homology group $H_1(\RS\setminus \{x_i\}_{i=n+1}^{m+n})$ of the  Riemann surface $\RS$ punctured at zeros of $v$ is, in analogy 
to (\ref{dualcont2}), given by
 \be
s^*_\a= -b_\a\;,\;\;  s^*_{\a+g}= a_\a \hskip0.5cm \a=1,\dots,g\;; \hskip0.5cm
s^*_{2g+k-n-1}= c_k\;,\hskip0.5cm k=n+2,\dots,n+m\;;
\la{dualcont3}
\ee
where $c_k$ is a small positively oriented contour around zero $x_k$ ($k=n+1,\dots,n+m$).

The variational formulas for matrix of $b$-periods, basic holomorphic differentials, Bergman bidifferential and bidifferential 
$Q_v$
coincide with the formulas (\ref{varO2})- (\ref{varQv2}) in the  case of arbitrary meromorphic differentials, excluding the 
derivatives with respect to residues.

\subsection{Bergman tau-function}
 
The Bergman tau-function $\tau_B(\RS,v)$ on  $\Hgkd$ is defined  by the following system of equations:
\be
\f{\p\log \tau_B(\RS,v)}{\p z_i}= -\f{1}{2\pi i} \int_{s_i^*} Q_v\;,
\la{deftau1}
\ee
or,
$$
\f{\p\log\tau_B(\RS,v)}{\p z_i}= -\f{1}{2\pi i}\int_{s_i^*}\f{ B_{reg}(x,x)}{v(x)}
$$
where the regularization of $B(x,y)$ is made using the differential $v$:
\be
B_{reg}(x,x)=\left( B(x,y)-\f{v(x)v(y)}{(\int_{x}^y v)^2}\right)\Big|_{y=x}
\la{regB}
\ee

The system (\ref{deftau1}) can be alternatively written in terms of the differential of $\log\tau_B$:
\be
d\log\tau_B(\RS,v)=-\f{1}{2\pi i} \sum_{i=1}^{2g+m-1} \left(\int_{s_i^*} Q_v\right) d\left(\int_{s_i} v\right)\;;
\la{tau2def}
\ee
the closedness of the 1-form in the right-hand side of (\ref{tau2def})  as well as for the case of the spaces of 
holomorphic differentials or Hurwitz spaces \cite{IMRN},
is an immediate corollary of variational formula (\ref{varQv2}).

The second derivatives of $\tau_B$ can be computed using variational formula (\ref{varQv2}):
\be
\f{\p^2}{\p z_i \p z_j}\log\tau_B(\RS,v)=-\f{1}{2\pi i} \int_{s_i^*} \int_{s_j^*} \f{B(x,y)}{ v(x)v(y)}\;.
\la{secontauB}
\ee

Equations (\ref{tau2def}) can be integrated similarly to \cite{JDG,IMRN}.
Let us introduce a distinguished set of local parameters on $\RS$ determined by the differential $v$.
In a neighborhood of a point $x_i$  belonging to divisor $(v)=\sum_{i=1}^{m+n}k_i x_i$ (if $k_i>0$ then $x_i$ is a
zero of order $k_i$, if $k_i<0$ then $x_i$ is a pole of order $-k_i$) the distinguished 
 local parameter is 
defined by the following two conditions:
\be
v=d(\zeta_i^{k_i+1})\;, \hskip0.7cm \zeta_i(x_i)=0
\la{defdist}
\ee
In the case when $x_i$ is a zero, i.e. $k_i>0$, conditions (\ref{defdist}) define $\zeta_i$ uniquely:
\be
\zeta_i(x)=\left[\int_{x_i}^x v\right]^{1/(k_i+1)}\;,\hskip0.7cm k_i>0
\la{distin}
\ee
If $x_i$ is a pole i.e. $k_i\leq -2$,  conditions (\ref{defdist}) define $\zeta_i$ up to the choice of the initial integration point $x_0$, not
coinciding with any pole of $v$:
\be
\zeta_i(x)=\left[\int_{x_0}^x v\right]^{1/(k_i+1)}\;,\hskip0.7cm k_i\leq -2
\la{distin1}
\ee

However, it's important to notice that for two different choices $x_0$ and $\tilde{x}_0$ of the initial point of integration, we have for corresponding
 local parameters $\zeta_i$ and  $\tilde{\zeta}_i$ that
\be
\f{d\zeta_i(x)}{d\tilde{\zeta}_i(x)}\Big|_{x=x_i}=1
\la{equivdist}
\ee

Let us now introduce the following notations:
\be
E(x,x_i):= \left( E(x,y) [d\zeta_i(y)]^{1/2}\right)|_{y=x_i}\;,
\la{Exxi}
\ee
\be
E(x_i,x_j):= \left( E(x,y) [d\zeta_i(x)d\zeta_j(y)]^{1/2}\right)|_{x=x_i,\,y=x_j}\;.
\la{Exixj}
\ee
(we notice that, due to (\ref{equivdist}), these objects are independent of the choice of distinguished local parameters near poles of $v$).
Define two vectors $r,s\in \Z^g$ such that
\be
\Acal_x((v))+2K^x+ \O r+s=0
\la{fundom}
\ee

Then the expression for
the tau-function on the space of differentials of second kind  formally looks similar to the case when $v$ is holomorphic (formulas (3.24) and (3.25) of
\cite{JDG}):
\be
\tau_B(\RS,v)= \left({\mathcal F}^{4}\prod_{i,j=1\; i<j}^{m+n} E^{k_ik_j}(x_i, x_j)\right)^{1/6} e^{-\f{\pi i}{6}\langle r,\O r\rangle}\;,
\la{taumer2}
\ee
where 
\be
{\mathcal F}=\left[\f{v(x)}{\prod_{i=1}^{m+n} E^{k_i}(x,x_i)}\right]^{(g-1)/2} e^{-\pi i \langle r, K^x\rangle}\Ccal(x)
\la{defFcal2}
\ee
is independent of $x$. Here $\Ccal(x)$ is given by (1.17) of \cite{Fay92}; this is a multiplicative 
$g(1-g)/2$ - differential on $C$ which is expressible via derivatives of order $g$ of the theta-function at the
vector of Riemann constants.


Independence of ${\mathcal F}$ of $x$ follows from two observations. First, the combination of prime-forms in the
denominator of (\ref{defFcal2}) ``kills'' all poles and zeros of $v(x)$. Therefore, the ratio in front of $\Ccal(x)$ is holomorphic in $x$ in the fundamental domain of $\RS$. Since $\Ccal(x)$ is also holomorphic and non-vanishing in the fundamental domain, ${\mathcal F}$ as a whole is holomorphic and non-vanishing.
Moreover, a simple computation (using (\ref{fundom}))  shows that  the automorphy factors along $b$-cycles introduced by the prime-forms cancel against corresponding automorphy factors of $\Ccal(x)$. Finally, the tensor weight of ${\mathcal F}$ equals $0$  
($\Ccal(x)$ is a $g(1-g)/2$-differential, and the pre-factor in front of $\Ccal(x)$  is a $g(g-1)/2$ - differential since $\sum_{i=1}^{m+n} k_i=2g-2$).

For the case of Bergman tau-function on spaces of holomorphic differentials the dependence on the choice of differential for a fixed Riemann surface is given by a discrete version of Polyakov-Alvarez formula (see equation (4.66) of \cite{JDG}). 
Using the explicit expression for the tau-function (\ref{taumer2}), one can extend the discrete Polyakov formula to the case of meromorphic differentials.

\subsection{Baker-Akhiezer tau-function}

Here we are going to define {\it Baker-Akhiezer tau-function} which is a   combination  of the KP and  Bergman tau-functions.
 The Baker-Akhiezer tau-function is
 naturally associated to the 
Baker-Akhiezer spinor kernel.
The main guiding principle for our definition will be behavior of this tau-function under symplectic transformations, and possibility to define
it intrinsically in terms of Baker-Akhiezer spinor kernel as a section of a line bundle on moduli space of meromorphic differentials. 

\begin{definition}
For a Riemann surface $\RS$ of genus $g$, holomorphic line bundle $\chi$ of degree $g-1$ over $\RS$ and meromorphic differential $v$ with divisor type 
$(k_1,\dots,k_{m+n})$ the Baker-Akhiezer tau-function $\tau_{BA}$ is defined by the formula
\be
\tau_{BA}(\RS,v,\chi)= e^{-\pi i \langle p_v, q_v\rangle} \tau_B^{-1/2} \th[p_v,q_v](0)
\la{taucomb}
\ee
where $\tau_B=\tau_B(\RS,v)$ is the Bergman tau-function.
\end{definition}

 Similarly to the Bergman tau-function, which satisfies 
variational formulas involving regularization of the Bergman bidifferential $B(x,y)$ near diagonal $y=x$,  variational formulas for the Baker-Akhiezer
tau-function
involve regularization of the Baker-Akhiezer kernel near diagonal. In both cases the regularization procedure 
is performed using the differential $v$.

Namely, let us introduce the following  holomorphic Abelian differential:
\be
w(x)= S^{(v)}_{reg}(x,x)= \lim_{y\to x} \left( S^{(v)}(x,y)-\f{\sqrt{v(x)}\sqrt{v(y)}}{\int_x^y v}e^{\int_x^y v}\right)\;,
\la{defwx}
\ee
and the following meromorphic quadratic differential:
\be
\phi(x)= \left[S^{(v)}_{reg}(x,x)\right]^2 - \left[(S^{(v)})^2\right]_{reg}(x,x)\;,
\la{defqx}
\ee
where
$$
[(S^{(v)})^2]_{reg}(x,x)=  \lim_{y\to x}\left(S^{(v)}(x,y)S^{(v)}(y,x) +\f{v(x)v(y)}{(\int_x^y v)^2}\right)\;.
$$
The formulas (\ref{defwx}) and (\ref{defqx}) define $w(x)$ and $\phi(x)$ intrinsically in terms of the Baker-Akhiezer spinor kernel.

Using the  relationship between the BA kernel and the Szeg\"o kernel, we can express $w(x)$ and $\phi(x)$ in terms of coefficients of asymptotics of
Szeg\"o kernel near diagonal given by (\ref{a0}) and (\ref{a1}). Namely, if in (\ref{asdia}),   (\ref{a0}) and (\ref{a1}) we use the local parameter 
$\xi(x)$ such that $d\xi(x)= v(x)$, we get
\be
w(x)=\sum_{\a=1}^g \p_\a \log\th[p_v,q_v](0)\; u_\a(x)
\la{wx1}
\ee
where, as before, $\p_\a$ stands for derivative of theta-function with respect to its argument $z_\a$.

The quadratic differential $\phi(x)$ can be  rewritten using relation (\ref{SSB}) between the Szeg\"o kernel and $B(x,y)$, and the 
regularization of the Bergman bidifferential near diagonal given by (\ref{defQv}), (\ref{defQv1}):
\be
\phi(x)=\f{1}{6}(S_B-S_v) + \f{1}{\th[p_v,q_v](0)}\sum_{\a,\b=1}^g \p_\a\p_\b\th[p_v,q_v](0)\; u_\a(x) u_\b(x)
\la{qx1}
\ee
where $S_B$ is the Bergman projective connection, and $S_v$ is the projective connection given by (\ref{defSv}).

Denote by $N$ the least common multiple of $k_1+1,\dots,k_{m+n}+1$.
The key property of the Baker-Akhiezer tau-function is given in the following proposition.

\begin{proposition}\la{modinvBA}
The $24N$th power of the Baker-Akhiezer tau-function, $(\tau_{BA})^{24N}$, is independent of the choice  of canonical basis of cycles on $\RS$.
\end{proposition}
The {\it proof} follows from the formula (\ref{transdtau2}) below describing the transformation of the Bergman tau-function under change of 
canonical basis of cycles and transformation law (\ref{transth}) of the modified theta-function. 

$\Box$

Derivatives of the Baker-Akhiezer tau-function with respect to homological coordinates, assuming holonomies $p,q$ of the line bundle $\chi$  remain fixed,
 are given by the following proposition:

\begin{proposition}
The tau-function $\tau_{BA}(\RS,v,\chi)$ (\ref{taucomb}) satisfies the following variational formulas on the space $\Hgkd$:
\be
\f{\p}{\p z_i}\log \tau_{BA}=\f{1}{4\pi i} \oint_{s_i^*}  \f{\phi}{v} +\f{1}{2\pi i} \oint_{\mu(s_i^*)} (w+\f{v}{2})+\f{1}{4\pi i}
\log {\rm Hol}(\chi,\mu(s_i^*))\;,
\la{vartauh} 
\ee
$i=1,\dots,2g+m-1$, where $\mu$ is the natural projection from relative to absolute homology groups defined before Theorem \ref{BAvar}; holomorphic abelian differential $w$ and
meromorphic quadratic differential $\phi$ are expressed in terms of Baker-Akhiezer kernel via (\ref{defwx}), (\ref{defqx}). 
\end{proposition}

{\it Proof.} The multiplier $\th[p_v,q_v](0)$ of the tau-function $\tau_{BA}$ (\ref{taucomb}) depends on homological coordinates $z_i$ in two ways: via the matrix of $b$-periods $\Omega$ and via characteristic vectors $p_v,q_v$.

 Partial derivatives  of  $\th[p_v,q_v](0)$  with respect to  $z_i$ which correspond to $z_i$-dependence of the matrix  $\Omega_{\a\b}$,
can be computed using heat equation and  variational formulas for the matrix of $b$-periods (\ref{varO2}). Combining the derivatives 
of  $\th[p_v,q_v](0)$ with respect to $\Omega_{\a\b}$ with derivatives of the  Bergman tau-function  with respect to $z_i$, we get the 
first integral in the r.h.s. of (\ref{vartauh}).

  The second and third  terms in (\ref{vartauh}) correspond to dependence of the combination  
$$e^{-\pi i \langle p_v, q_v\rangle} \th[p_v,q_v](0)$$ on absolute periods of $v$ through the characteristic 
vectors $p_v$ and $q_v$. In computation of these terms we use formulas (\ref{modthp}) and (\ref{modthq}) for derivatives of 
the modified theta-function $\tilde{\th}$. Recall that holonomies of $\chi$ with respect to basic cycles are related to $p$ and $q$ via
(\ref{defholon}).

$\Box$


\begin{remark}\rm
The Bergman tau-functions $\tau_B$ and the Baker-Akhiezer tau-function $\tau_{BA}$ admit a natural interpretation from the point of view of
conformal field theory on Riemann surfaces, as well as from the point of view of the theory of determinants of Cauchy-Riemann operators.
There exist numerous references on these subjects; we mention here \cite{Knizhnik,VerVer,AG}  and the fundamental Quillen's paper \cite{Quillen}.

From the point of view of conformal field theory on Riemann surfaces (see \cite{Knizhnik,VerVer,AG}), the Bergman tau-function $\tau_B$ can be interpreted 
as chiral partition function of free bosonic theory corresponding to the system of local coordinates on $\RS$ defined by the differential $v$.
From the point of view of Quillen's theory   $\tau_B$ can be interpreted as a section of the determinant line bundle  corresponding to the operator $\bar{\p_0}$ acting 
on functions $f(x)$ with essential singularities on $\RS$ at poles of $v$ such that $f(x)e^{-\int^x v}$ is non-singular at these poles. 

On the other hand, the Baker-Akhiezer tau-function $\tau_{BA}$ is, from the point of view of CFT, the chiral partition function of the  system of free ``twisted fermions'' on $\RS$. From the point of view of Quillen's theory   $\tau_{BA}$  is a section of the determinant line bundle  corresponding to the operator $\bar{\p_\chi}$ acting 
on sections  $\psi$ of $\chi$  which have essential singularities on $\RS$ at poles of $v$ such that $\psi(x)e^{-\int^x v}$ is non-singular at these poles.

 Under  such interpretation the relation (\ref{taucomb}) 
can be considered as a version of  the ``bosonization''
formula from \cite{Knizhnik,VerVer,AG}.  

\end{remark}

\subsection{Tau-functions as holomorphic sections of  line bundles on moduli space of  differentials of second kind}

Let us consider the transformation of the Bergman tau-function under symplectic change of the canonical basis of cycles on $\RS$.
Let us look first at the definition (\ref{tau2def}). A change of basis $\{s_i\}$ in  $H_1(\RS;\{x_i\}_{i=n+1}^{m+n})$ does not change the
right-hand side of  (\ref{tau2def}) since the basis of  dual cycles $\{s_i^*\}$ is transformed by the inverse matrix. 

However, the differential $Q_v$ transforms  under change of normalization of canonical basis of cycles according to
(\ref{transQv}), which implies the following transformation of $d\log \tau_B(\RS,v)$ under symplectic transformation $\sigma$:
\be
d\log \tau_B^\sigma(\RS,v)=d\log \tau_B(\RS,v)+\sum_{\a,\b=1}^g  [(c\O+d)^{-1} c]_{\a\b} \left(\int_{s_i^*}\f{u_\a(x) u_\b(x)}{v(x)}\right)d \left(\int_{s_i} v \right)
\la{transdtau}
\ee 
which, taking into account variational formula (\ref{varO2}) for the matrix of $b$-periods, can be written as follows:
\be
d\log\tau_B^\sigma(\RS,v)=d\log\tau_B(\RS,v)+ d \log [{\rm det} (c\O+d)]\;
\la{transdtau1}
\ee 
therefore, 
\be
\tau_B^\sigma(\RS,v)= \gamma(\sigma)[{\rm det} (c\O+d)] \tau_B(\RS,v)
\la{transdtau2}
\ee
where $ \gamma(\sigma)$ is a moduli-independent character of $Sp(2g,\Z)$.

Some information about $ \gamma(\sigma)$ can be deduced from the explicit formula (\ref{taumer2}). The only ambiguity in the formula (\ref{taumer2}) is due to
the root of sixth degree in (\ref{taumer2}), the roots of unity in the definition of the distinguished local parameters (\ref{distin}) and the square root ambiguity in the 
definitions (\ref{Exxi}) and (\ref{Exixj}).

Let us now introduce the least common multiple $N$ of the numbers $k_i+1$ (recall that $(v)=\sum_{i=1}^{n+m} k_i x_i$):
\be
N:= LCM (k_1+1,\dots,k_{m+n}+1)
\la{lcm}
\ee
 
Then  from the explicit formula (\ref{taumer2}) for the tau-function we see that the $12N$th degree of  the tau-function is free from the ambiguity related to the choice of the branches of all of the roots; therefore,$ \gamma^{12N}(\sigma)=1$.

We arrive at the following transformation law of $\tau_B^{12N}$:
\be
[\tau_B^\sigma(\RS,v)]^{12N}= [{\rm det} (c\O+d)]^{12N} \tau_B^{12N}(\RS,v)
\la{transdtau3}
\ee
 
Denoting by $\lambda$ the (pullback from the moduli space of Riemann surfaces of genus $g$) Hodge class on
 $\Hgkd $, we get the following
\begin{proposition}\la{tausec1}
The $12N$th power of the Bergman tau-function $\tau_B$ is a holomorphic non-vanishing section of $12N$th power of the Hodge 
line bundle $\lambda^{12N}$ on $\Hgkd $.
\end{proposition}

The holomorphy and non-vanishing of $\tau_B$ in the space  $\Hgkd $ follows either from the 
definition (\ref{tau2def}), or from the explicit formula (\ref{taumer2}).

The proposition \ref{tausec1} in particular implies that the $12N$th power of the Hodge line bundle, $\lambda^{12N}$, is isomorphic to
the  trivial line bundle  on the space $\Hgkd $. The isomorphism between  $\lambda^{12N}$ 
and the trivial line bundle is given by $\tau_B^{12N}$.

Let us now consider the projectivization  $\Hgkp $ of the space
 $\Hgkd$: on  $\Hgkp$ the pairs $(C,v)$ and
$(C,\kappa v)$ (where $\kappa$ is a constant independent of a point of a Riemann surface) are identified.
Denote by $L$ the tautological line bundle associated with the projection $\Hgkd$ $\to$
 $\Hgkp$.

In analogy to the formula (3.9) of \cite{MRL} one can easily verify (either using the definition (\ref{tau2def}) 
or the explicit expression
(\ref{taumer2}))  that the tau-function $\tau_B(\RS,v)$ transforms as follows under the rescaling of the differential $v$:
\be
\tau_B(\RS,\epsilon v)= \epsilon^{\f{1}{12}\sum_{i=1}^{m+n}\f{k_i(k_i+2)}{k_i+1}}\,\tau_B(\RS, v)
\la{homogen}
\ee

Combining this rescaling property with the Proposition \ref{tausec1}, we get the following characterization of $\tau_B$ as a
section of a line bundle over  $\Hgkd$:
\begin{proposition}\la{tausec2}
The $12N$th power of the Bergman tau-function $\tau_B$ is a holomorphic non-vanishing section of  
the line bundle $\lambda^{12N}\otimes L^{N\sum_{i=1}^{m+n}\f{k_i(k_i+2)}{k_i+1}}$ on $\Hgkp$.
\end{proposition}

Denote the first Chern class of the tautological line bundle by $\psi$: $\psi=c_1(L)$. Then Proposition \ref{tausec2} implies the following relation  in the rational Picard group ${\rm Pic}(\Hgkp)$
between the Hodge class $\l$ and the tautological class $\psi$ on 
 $\Hgkp$:
\be
\l= \f{1}{12}\sum_{i=1}^{m+n}\f{k_i(k_i+2)}{k_i+1} \psi\;.
\la{hodgetav}
\ee

Consider now the Baker-Akhiezer tau-function $\tau_{BA}$ on $\Hgkp$. According to  (\ref{transth}) and proposition \ref{tausec1},
$\tau_{BA}^{24N}$ is invariant under symplectic transformations of canonical basis of cycles. On the other hand, due to (\ref{homogen}), 
 $\tau_{BA}$ has the following homogeneity property:
\be
\tau_{BA}(\RS,\epsilon v)= \epsilon^{-\f{1}{24}\sum_{i=1}^{m+n}\f{k_i(k_i+2)}{k_i+1}}\tau_{BA}(\RS, v)
\la{homBA}
\ee

Therefore, $\tau_{BA}^{24N}$ is a holomorphic section (with zeros) of the line bundle $L^{-N\sum_{i=1}^{m+n}\f{k_i(k_i+2)}{k_i+1}}$.

The divisor of zeros $(\Theta^\chi)$  of $\tau_{BA}$ is defined by (\ref{defThetabig}); taking into account  the transformation law (\ref{transth}) of
the modified theta-function, we conclude that in the Picard group of  $\Hgkp$  we have    $(\Theta^\chi)=-\f{1}{2}\lambda$. Combining this with (\ref{hodgetav}), we get the following expression for the class of divisor $(\Theta^\chi)$ via the tautological class:
\be
(\Theta^\chi)=-\f{1}{24}\sum_{i=1}^{m+n}\f{k_i(k_i+2)}{k_i+1} \psi
\la{thchipsi}
\ee

In \cite{MRL,Advances} the Bergman tau-function was studied on compactifications of Hurwitz spaces and 
spaces of holomorphic differentials, which allowed to express the Hodge class on compactifications of these spaces in terms of boundary divisors (and the tautological class in the case of spaces of holomorphic differentials).

We don't perform similar analysis in the more general case of spaces of differentials of second kind considered here, since we are not aware of existence of a suitable compactification of these spaces.

\subsection{Special cases: Hurwitz spaces and spaces of holomorphic differentials}

The Hurwitz spaces are subspaces of spaces of differentials of second kind 
defined by vanishing of all absolute periods of the differential $v$; in that case the differential $v$ is exact: $v=df$ where $f$ is a meromorphic function on $\RS$. Denote the Hurwitz subspace of $\Hgkd$ defined
by vanishing of all absolute periods of the differential $v$: $\oint_{a_\a}v=\oint_{b_\a}v=0$ by
$\Hurgk$. In genus zero case, when the absolute periods are absent, the spaces
 $\Hgkd$ and $\Hurgk$ coincide; these are simple spaces of rational
 functions with fixed multiplicities of poles and critical points. The Bergman tau-function on spaces of rational functions was studied in 
detail in \cite{KokStr,Advances}.

On the space $\Hurgk$  only relative periods of the differential $v$ don't vanish: $\int_{x_1}^{x_i}df= f(x_i)-f(x_1)$, $i=2,\dots,m$. These are nothing but critical values $z_2,\dots,z_m$ of the function $f$ (the first critical value
$z_1$ in our conventions coincides with $0$); in other words, the zeros of $v=df$ are ramification points of the covering 
$f\;:\;\RS\to\C$ defined by function $f$; values of the function $f$ at these  points: $z_1=0, z_2,\dots, z_m$  are the branch points of the covering.

The transformation laws for the tau-function under the change of canonical basis on $\RS$ (\ref{transdtau3}) and under the rescaling of 
differential $v$ (\ref{homogen}) remain if $\tau_B$ is restricted on Hurwitz space  $\Hurgk$ from
$\Hgkd$. Therefore, propositions (\ref{tausec1}) and (\ref{tausec2}) remain in force for the case of the Hurwitz spaces. Actually, in the case of Hurwitz spaces one can say much more about geometrical meaning of $\tau_B$ 
due to existence of natural compactification of Hurwitz spaces given by spaces of admissible covers of Harris and Mumford \cite{HarMum}. 

First (see \cite{Advances} for details), in the case of Hurwitz spaces it is natural to introduce a stronger equivalence relation than in general case: instead of assuming that the pair $(\RS_1,v_1)$ is equivalent to the pair  $(\RS_2,v_2)$
if $\RS_1=\RS_2$ and $v_1= \kappa v_2$ for a constant $\kappa$, one can introduce a stronger equivalence relation: 
 $(\RS_1, df_1)$ and $(\RS_2, df_2)$ are equivalent if $\RS_1=\RS_2$ and $f_2$ is obtained by a M\"obius transformation from $f_1$.

This equivalence relation is most natural in the generic case: if we assume that all poles of $f$ and 
all zeros of $df$ are simple. That was the case treated in details in \cite{Advances}. 

Another important property of the Hurwitz subspaces is that the square of the  Vandermonde determinant of the critical values 
$$
V^2=\prod_{i<j} (z_i-z_j)^2
$$
for $i,j=1,\dots,m$, is a well-defined function on $\Hurgk$ (obviously, this is not the case if some absolute periods don't vanish).

Let us now restrict ourselves to the generic stratum of the Hurwitz space, assuming that all poles and all critical points
of the function $f$ are simple (i.e. all zeros of $v=df$ are simple, and all poles of $v$ have multiplicity two).

Denote such space by $\Hur_{g,n}$, where $n$ now equals the degree of the meromorphic function $f$.

Then, as was shown in \cite{Advances}, the following combination of the Bergman tau-function on the space  $\Hur_{g,n}$ and
the Vandermonde determinant $V$:
\be
\hat{\eta}= \frac{\tau_B^{24(m-1)}}{V^6} 
\la{etah}
\ee
is invariant under M\"obius transformation of the function $f$.

Let us denote by $\Hur^0_{g,n}$ the quotient of  $\Hur_{g,n}$ by the M\"obius equivalence relation: 
the point $(\RS_1,df_1)$ is equivalent to the point  $(\RS_2,df_2)$ if $\RS_1=\RS_2$ and $f_2=(a f_1+b)(c f_1 +d)^{-1}$ for some
$a,b,c,d\in \C$. 

On  $\Hur^0_{g,n}$ the combination $\hat{\eta}$ is a holomorphic non-vanishing 
section of $\lambda^{24(m-1)}$; thus on  $\Hur_{g,n}^0$ the 
line bundle  $\lambda^{24(m-1)}$ is isomorphic to the trivial line bundle.

The space  ${\Hur}^0_{g,n}$ admits a natural compactification  ${\bHur}^0_{g,n}$, called the space of admissible covers
\cite{HarMum}. On various components of the  boundary  ${\bHur}^0_{g,n}\setminus {\Hur}^0_{g,n}$ the section $\hat{\eta}$
has poles and zeros of easily computable degrees. This allows to express the $\lambda$-class in terms of boundary divisors
in the rational Picard group of the space  ${\bHur}^0_{g,n}$ (see the formula (3.13) of \cite{Advances}).

The Baker-Akhiezer tau-function in the case of Hurwitz spaces has the form:
\be
\tau_{BA}= e^{-\pi i \langle p, q\rangle}\tau_B^{-1/2} \th_{pq}(0) 
\la{BAHur}
\ee
(we have $p_v=p$ and $q_v=q$ since all absolute periods of $v$ vanish); the factor $e^{-\pi i \langle p, q\rangle}$ then does not depend on any moduli coordinates. 

The tau-function (\ref{BAHur}) is the Jimbo-Miwa tau-function of Riemann-Hilbert problems with quasi-permutation monodromy matrices (see \cite{Birk,Annalen}). The divisor of zeros of
$\tau_{BA}$ in the context  of the Riemann-Hilbert problems is called the Malgrange divisor $({\mathcal M})$; it consists of configurations of poles of the RH problem where the
latter becomes unsolvable. The previous analysis shows that on the open part of the Hurwitz space the class of Malgrange divisor is expressed via the Hodge class 
as $({\mathcal M})=-\l/2$, and can be further expressed via tautological class using (\ref{hodgetav}). However, it seems hard to write a nice universal expression for
$({\mathcal M})$ on the compactified Hurwitz space (i.e. space of admissible covers) due to complicated behavior of $\th_{pq}(0)$ near different boundary components.

Spaces of holomorphic differentials correspond to a  special case of the general construction presented in this section,
corresponding to $n=0$.
The geometrical properties of the Bergman tau-function on the   space holomorphic differentials with simple zeros were treated in detail in \cite{MRL}.
The space  of holomorphic differentials with simple zeros admits a natural compactification given by the Hodge vector bundle over Deligne-Mumford 
compactification of the moduli space of Riemann surfaces.
In this compactification the boundary divisors  are the (pullbacks of) the boundary 
divisors $D_0,\dots, D_{[g/2]}$ of  Deligne-Mumford compactification. An additional boundary component $D_{deg}$ corresponds to 
the space  of differentials having one zero of multiplicity 2 and other zeros simple.
Computing the asymptotics of the tau-function near these boundary components one can express the Hodge class in terms
of the boundary divisors and the first Chern class of the tautological line bundle 
(see formula (4.6) from \cite{MRL}).

\subsection{Genus one: Bergman tau-function and modular forms}

In genus zero spaces of meromorphic differentials coincide with spaces of rational functions; for this case the Bergman tau-function was studied in detail in \cite{IMRN,Advances}.
New non-trivial examples first appear in genus one. 

Since  holomorphic differential $v$ on Riemann surface of genus one is unique up to a constant, the corresponding 
Bergman tau-function depends 
only on $a-$ and $b$-periods of $v$ (call them $A$ and $B$); it coincides with the square of the corresponding 
Dedekind's eta-function (formula (1.6) of \cite{JDG}):
\be
\tau_B(\RS,v)=\eta^2(B/A)
\la{taueta}
\ee
The 24th power of this tau-function, $\tau_B^{24}=\eta^{48}$, is a modular form of weight $24$ and level 1. 
Essentially this is  the only interesting genus 1  example if we restrict ourselves by holomorphic differential $v$.

However, once we  allow $v$ to have singularities,  we can get other examples of modular form.
Namely, consider the space $\Hcal_1(-k,k)$, which consists of pairs $(\RS,v)$, where $\RS$ is the Riemann surface of genus one, 
and $v$ is a meromorphic differential on $\RS$ with one pole of order $k$ and one zero of order $k$.
Then $v=h^k$, where $h$ is a meromorphic section of a ``$k$-spin'' line bundle $\chi$ over $\RS$ (the $k$-spin line bundle 
is a root of $k$th degree from the canonical line bundle, i.e. $\chi^k$ is the canonical line bundle over $\RS$).
The ``$k$-spinor'' $h$ has simple pole at $x_1$ and  simple zero at $x_2$;  
the divisor $k(x_2-x_1)$ is trivial.
Choose on $\RS$ the basic cycles $a$ and $b$,  introduce the corresponding normalized holomorphic differential $u$ and  
denote its $b$-period by $\mu$.
Then $\int_{x_1}^{x_2} u   = \f{p}{k} \mu +\f{q}{k} $ for some integers $p$ and $q$ which 
depend on the choice of the $k$-spinor line bundle $\chi$ and the choice of canonical cycles $(a,b)$.
Let
$$
\a:=\f{p}{k}\;,\hskip0.7cm \b:=\f{q}{k}
$$

Using the holomorphic differential $u$, let us introduce on $\RS$ the coordinate $\lambda$:
\be
\lambda(x)=\int_{x_2}^x u
\ee
 (i.e. $u=d\l$).
Then $\l(x_1)=0$ and $\l(x_2)=\a \mu +\b$ and the differential $v$ can be written as follows:
\be
v=  C \left(\frac{\th[\f{1}{2}-\a,\f{1}{2}-\b](\l)}{\th[\f{1}{2},\f{1}{2}](\l)}\right)^k d\l
\la{vfirst}
\ee
where $\th[1/2,1/2]=\th_1$ is the odd Jacobi theta-function;  $C$ is a constant which may depend on the module of the curve.

To get an explicit formula for the Bergman tau-function from the general expression (\ref{taumer2}) one needs to compute various ingredients of  (\ref{taumer2}). 
In genus one case the  Riemann constant $K^x$ is independent of $x$ and given by $K^x=(\mu+1)/2$.
Relation (\ref{fundom}) allows to express   integers $r$ and $s$       via $\a$ and $\b$. Namely, we have
$$
\Acal_x((v))= k(\a \mu +\b)\;;
$$
therefore, the relation $\Acal_x((v))+2K^x+r\mu+s=0$ implies
$r=-(k\a+1)$, $s=-(k\b+1)$.

Then 
$$
{\mathcal F}= e^{\f{\pi i}{2}(\mu +1)(k\a+1)} \th'(\f{\mu+1}{2})
$$
and
\be
\tau_B= e^{\f{\pi i}{6}(1-k^2\a^2)\mu} \left[\th'\left(\f{\mu+1}{2}\right)\right]^{2/3} [E(x_2,x_1)]^{-k^2/6}\;.
\la{tauE}
\ee

The prime-form $E(\l_1,\l_2)$ is given by:  
\be
E(\l_1,\l_2)=\f{\th_1(\l_1-\l_2)}{\th_1'(0) \sqrt{d\l_1}\sqrt{d\l_2}}\;;
\ee 
thus
\be
E(x_2,x_1)=\f{\th_1(\a \mu +\b)}{\th_1'(0)}
\left[\f{d\zeta_1(x)}{d\l(x)}\Big|_{x=x_1} \f{d\zeta_2(x)}{d\l(x)}\Big|_{x=x_2} \right]^{1/2}\;,
\la{Exx}\ee
where $\zeta_1$ and $\zeta_2$ are distinguished local parameters near the pole $x_1$ ($\l=0$) and the zero $x_2$ ($\l=\a \mu +\b$),
respectively.

We have $\zeta_1(x)= \left[\int^x v\right]^{-1/(k-1)}$ where the initial point of integration can be chosen arbitrarily. Thus  near $x_1$ we have
$$
v\sim C \left[\f{\th[\f{1}{2}-\a,\f{1}{2}-\b](0)}{\th_1'(0)}\right]^k \f{d\l}{\l^k}\;,
$$
and, up to a moduli-independent constant:
\be
\f{d\zeta_1(x)}{d\l(x)}\Big|_{x=x_1}= C^{-1/(k-1)}\left[\f{\th[\f{1}{2}-\a,\f{1}{2}-\b](0)}{\th_1'(0)}\right]^{-k/(k-1)}\;.
\la{zeta0}
\ee

For the distinguished local parameter near the zero $x_2$ we have $\zeta_2(x)= \left[\int_{x_2}^x v\right]^{1/(k+1)}$. 

From the definition of theta-function with characteristics one gets the following relation:
\be
\th\left[\f{1}{2}-\a,\f{1}{2}-\b\right](\lambda)=e^{ \pi i \mu \a^2-2\pi i \a(\lambda-\b+{1}/{2})} \th\left[\f{1}{2},\f{1}{2}\right](\lambda-\mu \a-\b)\;,
\la{thth}
\ee
which implies that, up to a moduli-independent constant,
\be
v\sim  C \left[\f{e^{-\pi i \mu \a^2}\th_1'(0)}{\th_1(\a\mu+\b)}\right]^k (\l-\a\mu-\b)^k d\l\;,
\ee
as $\l\to \a\mu+\b$.
Therefore,
$$
\f{d\zeta_2(x)}{d\l(x)}\Big|_{x=x_2}= C^{1/k+1} \left[\f{e^{-\pi i \mu \a^2}\th_1'(0)}{\th_1(\a\mu+\b)}\right]^{k/(k+1)}\;,
$$
or, using (\ref{thth}),
\be
\f{d\zeta_2(x)}{d\l(x)}\Big|_{x=x_2}= C^{1/k+1} \left[\f{\th_1'(0)}{\th[{1}/{2}-\a,{1}/{2}-\b](0)}\right]^{k/(k+1)}\;.
\la{zeta1}
\ee

Combining (\ref{Exx}) with (\ref{zeta0}) and (\ref{zeta1}), we get, up to a moduli-independent constant:
\be
E(x_2,x_1)=e^{-\pi i \mu \a^2} \left[\f{\th_1'(0)}{C\, \th[{1}/{2}-\a,{1}/{2}-\b](0)}\right]^{1/(k^2-1)}\;.
\la{Exx1}
\ee
Substituting (\ref{Exx1}) into  (\ref{tauE})
we get
\be
\tau_B= C^{\f{k^2}{6(k^2-1)}}\left\{\th_1'(0)\right\}^{\f{2}{3}+\f{1}{6}\f{k^2}{1-k^2}} 
\left\{\th\left[\f{1}{2}-\a,\f{1}{2}-\b\right](0)\right\}^
{\f{k^2}{6(k^2-1)}}\;.
\la{taumod}
\ee

The natural choice of the constant $C$ is  $C=[\th_1'(0)]^{2/3}$, which makes the differential $v$ moduli-invariant up to  root of unity; then (\ref{taumod}) turns into
\be
\tau_B=  \left\{\th_1'(0)\right\}^{\f{2}{3}+\f{1}{18}\f{k^2}{1-k^2}} \left\{\th\left[\f{1}{2}-\f{p}{k},\f{1}{2}-\f{q}{k}\right](0)\right\}^{\f{k^2}{6(k^2-1)}}
\la{taumod1}
\ee
For $k$ even the expression $(\tau_B)^{9(k^2-1)}$ is a modular form of level $k$ and weight $9(k^2-1)$.
For $k$ odd  the expression $(\tau_B)^{18(k^2-1)}$ is a modular form of level $2k$ and weight $18(k^2-1)$.
There exist several known cases when tau-functions arising in various 
contexts transform as modular forms \cite{Hitchin,Takhtajan,BabKor,HarnadMckay}; 
here we get new examples of this kind.

\section{Bergman tau-functions on  spaces of differentials of third kind}

\la{sectionTHIRD}

\subsection{Definition and basic properties}

Here we briefly outline the construction and properties of the Bergman tau-function on moduli spaces 
of differentials of third kind.
The study of Bergman tau-function on spaces of differentials of third kind is important due to  possible 
interpretation of this tau-function as chiral partition function of free bosons on Mandelstam diagrams \cite{DH}.

However, presence of simple poles with non-vanishing residues makes the definition of the Bergman 
tau-function more difficult. To explain the source of these difficulties we notice first that all variational formulas
(\ref{varO2}) - (\ref{varQv2}) are valid for differentials with non-vanishing residues (irrespectively of the order of the poles), as well as for differentials
without residues.  In particular, derivative with respect to a  residue of $v$ at $x_k$ involves in the right-hand side an integral over contour which is dual to the small circle around the pole; this dual contour (denoted by $-\tilde{l}_k$ in
(\ref{dualcont2}) connects a chosen ``first'' pole with pole number $k$. The r.h.s. of formulas 
(\ref{varO2}) - (\ref{varQv2}) are well-defined since all differentials ($u_a u_b/v$ etc) integrated along  $-\tilde{l}_k$
are non-singular (and, moreover, vanishing) at poles of $v$ i.e. at the endpoints of  $-\tilde{l}_k$.

A naive attempt  to extend the definition (\ref{deftau1}) of the Bergman tau-function from the space of
differentials of second kind to the space of differentials having simple poles with residues would be to choose
$\{s_i\}$ to be a full set of generators (\ref{defsmerom}) of the homology group (\ref{relhol1}); then the set of dual generators is given by (\ref{dualcont2}). However, this attempt fails since $Q_v$ (in contrast to differentials in the r.h.s. of variational formulas (\ref{varO2})-(\ref{varQv2})) has simple poles at the endpoints of
contours  $-\tilde{l}_k$ (i.e. at the simple poles of $v$).

Here we circumvent this problem by considering  all residues at the poles of $v$ constant, and not varying the tau-function with respect to them. This assumption can not be considered completely satisfactory, in particular, since it makes the
definition of the Bergman tau function more ambigous than in the case of differentials of second kind.
However, we follow this assumption here and study in detail the first non-trivial case of moduli space of rational
functions with 4 simple poles.

Following notations of Section \ref{Vfmd},  let differential $v$ have $n$ simple poles at
$x_1,\dots,x_n$
 with given residues $r_1,\dots,r_n$ ($r_1+\dots+r_n=0$)
and $m=2g-2+n$ simple zeros (we denote them by  $x_{n+1},\dots,x_{n+m}$).

Denote  moduli space of Riemann surfaces equipped with such differential  by $\Hmr$. 
Assuming that the Riemann surface $\RS$ is weakly-marked i.e. equipped with a chosen canonical basis of cycles, 
we get covering of the space $\Hmr$,
which we denote by $\Hmrt$.

To introduce homological coordinates on $\Hmr$ we consider relative homology group (\ref{relhol1})
and choose a set of generators (\ref{defsmerom}).  The homological coordinates $\oint_{c_{k+1}}v=2\pi i r_{k+1}$ are
going to be kept fixed; the remaining $2g+m-1$ generators will be denoted by
\be
s_\a= a_\a\;,\;\;  s_{\a+g}= b_\a \hskip0.5cm \a=1,\dots,g\;; \hskip0.5cm
s_{2g+k-1}= l_{n+k}\;,\hskip0.5cm k=2,\dots,m\;;
\la{defsi}
\ee
where, as before,
$l_k$ is a contour $[x_{n+1},\,x_k]$ connecting the first chosen zero $x_{n+1}$ with zero $x_k$.

To define generators (\ref{defsi}) we introduce the fundamental polygon $\hat{\RS}$ by cutting $\RS$ along the basic cycles $(a_i,b_i)$,
and introduce inside of $\hat{\RS}$ a system of cuts (denote it by $L$) connecting, say,  the first pole $x_1$ with other poles $x_2,\dots, x_n$.
Let us now connect the first zero $x_{n+1}$ with other zeros $x_2,\dots,x_{n+m}$ by contours  $[x_{n+1},\,x_k]$ lying in $\hat{\RS}\setminus L$.

The differential $Q_v$ has poles of third order (generically with non-vanishing residues) at $x_{n+1},\dots,x_{n+m}$.
What makes the present case essentially different from the case of holomorphic differentials or differentials of the second kind is that $Q_v$ also has first order poles
at the poles $x_1,\dots,x_n$ of $v$ with corresponding residues $-1/(12r_1),\dots,-1/(12r_n)$.

 The set of ``dual'' cycles $s_i^*$  is now defined as follows:
\be
s^*_\a= -b_\a\;,\;\;  s^*_{\a+g}= a_\a \hskip0.5cm \a=1,\dots,g\;; \hskip0.5cm
s^*_{2g+k-1}= \tilde{c}_{n+k}\;,\hskip0.5cm k=2,\dots,m\;;
\la{defsist}
\ee
where $\tilde{c}_{n+k}$ are small circles around the zeros $x_{n+k}$. The dual cycles (\ref{defsist}) are uniquely determined by the choice of the
cycles (\ref{defsi}).


There are the following two sources of ambiguity in the construction of cycles $\{s_i\}$ and  $\{s_i^*\}$ relevant in 
the definition of Bergman tau-function.

\begin{itemize}
\item
The choice of the fundamental domain $\hat{\RS}$ and the system of cuts $L$ connecting poles of $v$.

\item
For fixed  fundamental domain $\hat{\RS}\setminus L$  the initial point of contours connecting zeros of $v$ can be put to
any zero $x_{n+j}$, instead of the first zero $x_{n+1}$. In the construction of the Bergman tau-function and variational formulas 
the same point $x_{n+j}$ should then be chosen as initial point of integration in defining the flat coordinate $z(x)$.

\end{itemize}

For a given choice of the contours $\{s_i\}$ and $\{s_i^*\}$ we  now introduce the following 1-form, which is defined  locally on the space  $\Hmr$:
\be
{\bf q}= \sum_{i=1}^{2g+m-1} \left(\int_{s_i^*} Q_v\right) d\left(\int_{s_i} v\right)
\la{defq1}
\ee
In contrast to the case of differentials of second kind, the form ${\bf q}$ is not globally well-defined on $\Hmr$.

Suppose that the  fundamental domain $\hat{\RS}$ and the system of cuts $L$ inside of $\RS$ are fixed. Let us change the
``first'' zero $x_{n+1}$ to another zero $x_{n+j}$ as the initial integration point. Denote the form ${\bf q}$ corresponding to this choice of 
homological coordinates by $\tilde{{\bf q}}$ (in the case of differentials of second kind we had $\tilde{{\bf q}}={\bf q}$). Now, since,
$\sum_{i=1}^{m} {\rm res}|_{x_{n+i}} Q_v = \f{1}{12}\sum_{i=1}^n \f{1}{r_i}$, it is easy to check that
\be
\tilde{{\bf q}}-{\bf q}=\left(\f{1}{12}\sum_{i=1}^n \f{1}{r_i}\right) d\left(\int_{x_{n+1}}^{x_{n+j}}v\right)\;.
\la{qqt}
\ee

The form ${\bf q}$ is closed, $d{\bf q}=0$, as a corollary of variational formulas (\ref{varQv2}); notice that the transformation (\ref{qqt}) is compatible with the
closedness of ${\bf q}$ since $\tilde{{\bf q}}-{\bf q}$ is  a closed form.

The potential of ${\bf q}$ is called the Bergman tau-function  $\tau_B(\RS,v)$ on moduli space of the differentials of 3rd kind
\be
d\log \tau_B(\RS,v)= -\f{1}{2\pi i} {\bf q}\;,
\la{Berg3}
\ee
or, in other words, $\tau_B(\RS,v)$ is defined by the following equations:
\be
\frac{\p}{\p\left(\int_{s_i} v\right)} \log\tau_B(\RS,v) = -\f{1}{2 \pi i} \int_{s_i^*} Q_v\;.
\la{deftau3}
\ee

According to (\ref{qqt}), under change of the choice of the ``first'' zero the Bergman tau-function transforms as follows:
\be
\tau_B\to \tau_B\exp\left\{\left(\sum_{i=1}^n \f{1}{12 r_i}\right)\int_{x_{n+1}}^{x_{n+j}}v\right\}\;.
\la{ttt}
\ee

The explicit formula for $\tau_B$  solving (\ref{deftau3}) formally again looks like (\ref{taumer2}), (\ref{defFcal2}).
However, distinguished local parameters near simple poles $x_i$ look as follows:
\be
\zeta_i(x)= {\rm exp}\left\{\f{1}{r_i}\int_{x_{n+1}}^x v\right\}\;,
\la{dist3}
\ee
where the integration paths between $x_{n+1}$ and the neighborhoods of poles $x_i$ are chosen inside of the 
fundamental domain $\hat{\RS}\setminus L$.

The degree with which the distinguished local parameter $\zeta_i(x)$ enters the formula (\ref{taumer2}) equals $-1/12$ (both products of
prime-forms in (\ref{taumer2}) contribute to that).
Therefore, the change of the initial integration point from $x_{n+1}$ to $x_{n+j}$ leads to multiplication of the explicit expression (\ref{taumer2})
with the factor
$$
\exp\left\{\left(\sum_{i=1}^n \f{1}{12 r_i}\right)\int_{x_{n+1}}^{x_{n+j}}v\right\}\;,
$$
in agreement with (\ref{ttt}).

Let us now discuss the dependence of $\tau_B$ on the choice of the fundamental polygon $\hat{\RS}$ and the system of cuts  $L$ between poles of $v$.
If the homology classes of $a$ and $b$-cycles are preserved, but positions of corresponding cuts forming the fundamental polygon $\hat{\RS}$ 
are changed (such that some of zeros or poles $x_{i}$ cross the geometrical positions of the cuts), the integration contours in
the definition of distinguished local parameters (\ref{dist3}) change by adding a combination of $a$ and $b$-cycles, as well as of small
cycles around poles of $v$. The integrals of $v$ around its poles are constants on our moduli space, therefore, up to a constant factor,
the change of the fundamental polygon leads to the following transformation of the tau-function:
\be
\tau_B\to \tau_B\exp\left\{ \f{1}{12}\sum_{i=1}^{2g}\sum_{j=1}^n \f{S_{ij}}{r_j} \int_{s_i} v\right\}\;,
\la{transtauR}
\ee
where $S_{ij}$ is an arbitrary matrix with integer coefficients.

Finally, the change of homology classes corresponding to $a$ and $b$-cycles leads to multiplication of $\tau_B$ with the factor ${\rm det} (c\O+d)$
as in (\ref{transdtau2}) (again up to  moduli-independent factor).

\begin{remark}\rm
The Bergman tau-function $\tau_B$ on the space of differentials of third kind can be used to define the partition function of
free bosons on Mandelstam diagrams \cite{DH}. Formally, the partition function for a given Riemann surface is given by 
$\left(\f{{\rm det}{\Delta}}{{\rm Area} (\RS)}\right)^{-1/2}$, where $\Delta$ is the Laplace operator in a given metric. In the case of Mandelstam diagram one uses the flat metric with conical 
singularities and cylindrical ends given by $|v|^2$ where $v$ is the differential of third kind with real residues, normalized by the condition that all of its $a$ and $b$-periods are imaginary. For such metric ${\rm det}{\Delta}$ is ill-defined, and ${\rm Area} (\RS)$
is infinite. However, using an analogy with the case of homolorphic differential \cite{JDG}, one can naturally define such partition function using the Bergman tau-function $\tau_B$ as follows:
\be
\f{{\rm det}{\Delta}}{{\rm Area} (\RS)}:={\rm const}\, |\tau_B(\RS,v)|^2 {\rm det}(\Im \Omega)\;.
\la{detDel}
\ee
We notice here that, although $\tau_B$ itself is defined on the moduli space only up to transformation (\ref{transtauR}), $|\tau_B|^2$
is uniquely defined if all $a$ and $b$- periods of $v$ are imaginary and all residues $r_i$ are real, which is the case on
the Mandelstam diagrams. For the rigorous spectral definition of ${\rm det}\Delta$, such
that (\ref{detDel}) becomes the {\it theorem}, not the definition, see \cite{KalKok}.

\end{remark}

\subsection{Four simple poles in genus zero}

Here we consider the simplest non-trivial case when $\RS$ is the Riemann sphere and the  differential $v$ has on $\RS$
four simple poles (which by a M\"obius transformation can be mapped to $0$, $1$, $t$ and $\infty$) with residues  
$r_1$, $r_2$, $r_3$ and $r_4$. Such differential has the form:
\be
v=\left(\frac{r_{1}}{x}+\frac{r_{2}}{x-1}+\frac{r_{3}}{x-t}\right)\,dx\;,
\la{v4poles}
\ee
with $r_{1}+r_{2}+r_{3}+r_{4}=0$ ($r_{4}:=r_{\infty}$).
The differential $v$ has two zeros which are solutions of the quadratic equation
\be
r_1(x-1)(x-t)+r_2 x(x-t)+ r_3x(x-1)=0\;,
\la{quadeq}
\ee
which are given by
\be
x_{1,2}=-\f{1}{2 r_4}(r_1+r_3 + (r_1+r_2) t \pm \sqrt{\Delta})\;,
\ee
where
$$
\Delta= ((r_1+r_2) t  + r_1 +r_3)^2 + 4 r_1 r_4 t\;.
$$
Equivalently,
$$
\Delta= (r_1+r_2)^2 (t-t_1)(t-t_2)\;,
$$
where we denote
\be
t_{1,2}=\frac{-(r_{1}r_{4}+r_{2}r_{3})\pm \sqrt{4\,r_{1}r_{2}r_{3}r_{4}}}{(r_{1}+r_{2})^{2}}\;.
\ee
The differential  $v$  (\ref{v4poles}) can be written as
$$
v=-r_4 \f{(x-x_1)(x-x_2)}{x(x-1)(x-t)}dx\;;
$$
if $t=t_1$ or $t=t_2$, zeros of $v$ coincide: $x_1=x_2$.

The moduli space of these differentials (with fixed residues $r_i$) is one-dimensional, it can  be parametrized by the position of the pole $t$;
in the generic case, when all $r_i$ are different, $t$ can be used as a global coordinate on the moduli space.

Let  us also introduce the following coordinate $\zeta$, which is defined on two-sheeted covering of $t$-plane:
\be
\zeta=\frac{2}{t_{1}-t_{2}}\left\{t-\frac{t_{1}+t_{2}}{2}+\sqrt{(t-t_{1})(t-t_{2})}\right\}\,;
\la{zetat}\ee
conversely, $t$ can be expressed via $\zeta$ as follows:
\be
t= \f{t_1-t_2}{4} (\zeta+\zeta^{-1})+\f{t_1+t_2}{2}\;.
\la{tzeta}
\ee
We have
$$
\f{d\zeta}{dt}=\f{4}{t_1-t_2}\f{\zeta^2}{\zeta^2-1} \;,
$$
and
\be
\zeta\f{dt}{d\zeta}=\sqrt{(t-t_1)(t-t_2)}\;.
\ee

Let us now assume that  $r_4=1$; this condition can be always achieved by rescaling of $v$.
Then the remaining residues can be parametrized by two constants 
$$   
p=\sqrt{\f{r_1}{r_2}}\;,\hskip0.7cm q=\sqrt{r_3}\;;
$$ 
then we have
$$
(r_1,r_2,r_3,r_4)=\left(-\f{p^2(1+q^2)}{1+p^2}, -\f{1+q^2}{1+p^2}, q^2,1\right)\;.
$$

Other objects can be expressed in term of $p$ and $q$ as follows:
\be
x_1=-\f{pq}{p^2+1}(\zeta-\f{p}{q})\;,\hskip0.7cm
x_2=-\f{pq}{p^2+1}(\zeta^{-1}-\f{p}{q})\;,\hskip0.7cm
t=\f{1+p^2}{p^2(1+q^2)} x_1 x_2\;,
\la{x12pq}
\ee
\be
x_{1,2}-1=-\f{pq}{p^2+1}(\zeta^{\pm 1}+\f{1}{pq})\;,
\la{x121}
\ee
\be
x_{1,2}-t=-\f{q^2}{(p^2+1)(q^2+1)}(\zeta^{\pm 1}-\f{p}{q})(\zeta^{\mp 1}+pq)\;.
\la{x12t}
\ee

 The homological coordinate can be chosen as
$$
z_1=\int_{x_1}^{x_2}v\;.
$$
It can be expressed as follows in terms of zeros $x_{1,2}$:
\be
z_1=\log\f{x_2^{r_1}(x_2-1)^{r_2}(x_2-t)^{r_3}}{x_1^{r_1}(x_1-1)^{r_2}(x_1-t)^{r_3}}\;,
\la{z1x12}
\ee
or, alternatively, in terms of $\zeta$ using (\ref{x12pq}), (\ref{x121}), (\ref{x12t}):
\be
z_1=\log\left\{\zeta^{2q^2}\left(\f{\zeta^{-1}-p/q}{\zeta-p/q}\right)^{\f{q^2-p^2}{1+p^2}}
\left(\f{\zeta^{-1}+1/pq}{\zeta+1/pq}\right)^{\f{q^2p^2-1}{1+p^2}}\right\}\;.
\la{z1pq}
\ee

An easy computation shows  that 
\be
\f{d z_1}{dt} = - \frac{\sqrt{\Delta}}{ t(t-1)}
\la{derDeltat}
\ee
and
\be
\f{d z_1}{d\zeta}=\f{p^2(q^2+1)(\zeta-\zeta^{-1})^2}{\zeta(\zeta-p/q)(\zeta^{-1}-p/q)(\zeta+pq)(\zeta^{-1}+pq)}\;.
\la{dz1dz}
\ee

The tau-function $\tau_B$ is now defined by only one equation
\be
\f{\p}{\p z_1}\log \tau_B= -{\rm res}|_{x_2} Q_v\;,
\la{deftau4p}
\ee
where
$$
Q_v=-\f{1}{6}
\left\{\f{1}{v}\left[\left(\f{v'}{v}\right)'
-\f{1}{2}\left(\f{v'}{v}\right)^2\right]\right\}\;.
$$

The derivative of $\log\tau_B$ with respect to $\zeta$ is obtained by combining (\ref{dz1dz}) and (\ref{deftau4p}).

The presence of simple zeros of $v$ makes the definition (\ref{deftau4p}) dependent on the choice of coordinate $z_1$.
Namely, let us now interchange the zeros $x_1$ and $x_2$; then we get an alternative homological coordinate $\tilde{z_1}=\int_{x_2}^{x_1} v=-z_1$.
Then instead of equation (\ref{deftau4p}) one gets the following equation, which defines a different tau-function $\tilde{\tau}_B$ via
\be
\f{\p}{\p \tilde{z}_1}\log \tilde{\tau}_B= -{\rm res}|_{x_1} Q_v\;.
\la{deftau4p1}
\ee
Since the sum of residues of $Q_v$ at $x_1$, $x_2$, $0$, $1$, $t$ and $\infty$ is zero, and the residue of $Q_v$ at a pole of $v$ equals 
$-1/(12r_i)$,
we have
$$
\f{\p}{\p \tilde{z}_1}\log \tilde{\tau}_B= {\rm res}|_{x_2} Q_v-\f{1}{12}\left(\f{1}{r_1}+\dots+\f{1}{r_4}\right)\;;
$$
thus, since $\tilde{z_1}=-z_1$,
\be
\f{\tilde{\tau}_B}{\tau_B}=\exp\left\{\f{z_1}{12}\left(\f{1}{r_1}+\dots+\f{1}{r_4}\right)\right\}\;,
\la{tauttau}
\ee
or
\be
\left(\f{\tilde{\tau}_B}{\tau_B}\right)^{12}=\left(\f{x_2^{r_1}(x_2-1)^{r_2}(x_2-t)^{r_3}}{x_1^{r_1}(x_1-1)^{r_2}(x_1-t)^{r_3}}\right)^{1/r_1+\dots+1/r_4}\;.
\la{tauttau1}
\ee

Since 
$$
\f{1}{r_1}+\dots+\f{1}{r_4}=\f{(p^2-q^2)(1-p^2q^2)}{p^2q^2(1+q^2)}\;,
$$
we have, taking into account (\ref{z1pq}):
\be
\f{\tilde{\tau}^{12}_B(\zeta)}{\tau^{12}_B(\zeta)}=C\;\zeta^{2\f{(p^2-q^2)(1-p^2q^2)}{p^2(1+q^2)}}
\left(\f{\zeta^{-1}-p/q}{\zeta-p/q}\right)^{\f{(q^2-p^2)^2(p^2q^2-1)}{(1+p^2)(1+q^2)}}
\left(\f{\zeta^{-1}+1/pq}{\zeta+1/pq}\right)^{\f{(q^2p^2-1)^2(q^2-p^2)}{p^2q^2(1+p^2)(1+q^2)}}\;,
\la{tauttau2}
\ee
where $C$ is an integration constant.

The straightforward integration of equation (\ref{deftau4p1})  leads to the following result:
\begin{proposition}
The Bergman tau-function on the space of differentials (\ref{v4poles}) is given by the following expression:
\be
\tau^{12}_B(\zeta)=
\zeta^{m_1} \left(\zeta^{-1}-\f{p}{q}\right)^{m_2} \left(\zeta-\f{p}{q}\right)^{m_3}
\left(\zeta^{-1}+\f{1}{pq}\right)^{m_4}\left(\zeta+\f{1}{pq}\right)^{m_5}(\zeta-\zeta^{-1})\;.
\la{Ber4poles}
\ee
where
$$
m_1=\f{(q^2-p^2)(1-p^2q^2)}{p^2(1+q^2)}\;,
$$
$$
m_2= 1+ \f{(q^2-p^2)^2}{p^2 q^2(p^2+1)(q^2+1)}\;,
$$
$$
m_3= 1+ \f{(q^2-p^2)^2}{(p^2+1)(q^2+1)}\;,
$$
$$
m_4= 1+ \f{(1-q^2p^2)^2}{q^2(p^2+1)(q^2+1)}\;,
$$
$$
m_5=1+ \f{(1-q^2p^2)^2}{p^2(p^2+1)(q^2+1)}\;.
$$
\end{proposition}

Comparing  (\ref{Ber4poles}) with the transformation (\ref{tauttau1}), we see that $\tilde{\tau}_B(\zeta)=\tau_B(\zeta^{-1})$,
i.e. tau-functions corresponding to two  different choices of the homological coordinate $z_1$  are simply different branches
of the tau-function   (\ref{Ber4poles}) (which correspond to variable $t$ belonging to different sheets).

In the  $\zeta$-plane the Bergman tau-function has 8 singular points.
The zeros at $\zeta=\pm 1$ corresponds to merging zeros of the differential $v$. Other six singularities correspond to
merging poles of $v$, when
 $t=0$, $t=1$ or $t=\infty$
(since $\zeta$-plane is two-fold covering of $t$-plane, these three values of $t$ correspond to six different values of  $\zeta$).

\subsection{Bergman tau-function on spaces of differentials with higher order poles with residues}

Surprisingly enough, the problems with invariant definition of the Bergman tau-function in the case of differentials of third kind
disappear if all poles of $v$ have order higher than 1, even though $v$ has non-trivial residues at these poles. 
Namely, consider the space  $\Hgk$ such that $k_i\neq -1$ for any $i$. 
Then  the differential $Q_v$ has zeros at all poles $x_1,\dots, x_n$ of $v$.  
As it was discussed in Section \ref{Vfmd}, the set of local homological coordinates is given by integrals (\ref{homcoormain})
 of $v$ over set of generators $\{s_i\}$ (\ref{defsmerom}) of the relative homology group (\ref{relhol1}). 
The dual generators $\{s_i^*\}$ of the dual relative homology group (\ref{defsstarmer}) are given by (\ref{dualcont2}).

Then the 1-form in the definition of the Bergman tau-function:
\be
d\log\tau_B(\RS,v)=-\f{1}{2\pi i} \sum_{i=1}^{2g-2+m+n} \left(\int_{s_i^*} Q_v\right) d\left(\int_{s_i} v\right)
\la{tauresid}
\ee
is well-defined on  $\Hgk$, in contrast to the case of when some of poles are of first order. This 1-form is closed as corollary of 
the variational formula (\ref{varQv2}), similarly to the case of differentials of second kind.

All the main properties of the Bergman tau-function on spaces of differentials of second kind extend to the case of differentials with residues:
the transformation law under the symplectic transformations (\ref{transdtau3}) as well as homogeneity property (\ref{homogen}). 
Therefore, Proposition \ref{tausec2}, which characterizes $\tau_B$ as a section of certain line bundle,  remains valid also  the case 
of differentials with residues.

As far as an explicit formula for the tau-function is concerned,  expressions (\ref{taumer2}) and (\ref{defFcal2}) also remain valid.
The only modification which has to be done in these formulas is  the definition of distinguished local parameters (\ref{distin}). Namely,
these parameters remain the same in the case of zeros of $v$. However,  distinguished local parameters near poles $x_i$, $i=1,\dots, n$, should, instead of 
(\ref{defdist}),  be defined from transcendental equations:
$$
v=\left(\f{r_i}{\zeta_i} +\f{1-k_i}{\zeta_i^{k_i}}\right)d\zeta_i
$$
or
\be
r_i \ln \zeta_i +\f{1}{\zeta_i^{k_i-1}} =\int_{x_0}^x v
\la{defzetai}
\ee
In contrast to the case of differentials of second kind ($r_i=0$), when $\zeta_i$ can be easily expressed via antiderivative of $v$ via (\ref{distin1}),
for $r_i\neq 0$ we can not find a simple expression for $\zeta_i$ from (\ref{defzetai}). As well as in the case of differentials of second kind ($r_i=0$),
$\zeta_i$ depend on the choice of the initial point of integration $x_0$.
However, in analogy to (\ref{equivdist}), we can conclude again that if $\zeta_i$ and $\tilde{\zeta}_i$ are local parameters corresponding to two different initial points of integration $x_0$ and $\tilde{x}_0$, then
$$
\f{d\zeta_i(x)}{d\tilde{\zeta}_i(x)}\Big|_{x=x_i}=1\;.
$$
Therefore, the freedom of choosing the initial point $x_0$ does not influence the explicit formula (\ref{taumer2}), (\ref{defFcal2}) for $\tau_B$.

The necessity to solve the transcendental equations (\ref{defzetai}) for distinguished local parameters makes the formula (\ref{taumer2})  somehow ``less explicit'' in the case of non-vanishing residues in comparison with the case of spaces of differentials of second kind.

Finally, we notice that one can view the spaces of differentials with higher order poles and non-vanishing residues as ``master'' moduli spaces.
Then the spaces of differentials of second kind or differentials with simple poles can be viewed as different boundary strata of this
master moduli space. From the point of view of Bergman tau-function the boundary stratum of differentials of second kind is the ``good'' one:
the Bergman tau-function there is invariantly-defined as a section of the same line bundle as on the master moduli space.
On the other hand, the boundary stratum consisting of differentials of third kind does not possess the same good properties: as we discussed above, the Bergman tau-function there
does not admit  an equally invariant definition.

{\bf Acknowledgements.} The work of DK was  partially supported by NSERC, FQRNT and Concordia University Research Chair grants.
We thank M.Bertola, A.Kokotov and P.Zograf for discussions. We thank the referee for careful reading of the manuscript and
useful suggestions.


\begin{thebibliography}{}
\bibitem{Abikoff} Abikoff, W., {\it The real analytic theory of Teichm\"uller space. Lecture Notes in Math.}, {\bf 820} Springer, Berlin, 144 p. (1980)

\bibitem{AG} Alvarez-Gaum\'e,L., Bost, J.-B., Moore, G., Nelson, Ph.,
  Vafa,C., {\it Bosonization on higher genus Riemann surfaces}, Commun.Math.Phys., 112,
503-552 (1987)

\bibitem{BabKor} Babich, M.,  Korotkin, D., 
{\it Self-dual $SU(2)$ invariant Einstein metrics and modular
dependence of theta-functions}, Letters in Mathematical Physics {\bf 46} 323-337 (1998)

\bibitem{BE} G.Borot, B.Eynard, {\it Geometry of spectral curves and all order dispersive integrable system}, 
arxiv: 1110.4939 (2011)

\bibitem{BisBZ} I.Biswas, Ben-Zvi, Theta Functions and Szeg\"o Kernels,  arXiv:math/0211441


\bibitem{Bobenko} Bobenko, A., {\it Helicoids with handles and Baker-Akhiezer spinors}, Math.Z., {\bf 229} 9-29 (1998)

\bibitem{Dubr} Dubrovin, B., Geometry of 2D topological field
theories, in: {\it Integrable systems and quantum groups. Proceedings,} Montecatini
Terme, 1993, pp. 120-348, Lecture Notes in Math., v.1620, Berlin:
Springer, 1996

\bibitem{Dubr1} Dubrovin, B., {\it  Painlev\'e transcendents in two-dimensional topological field theory. The Painlevé property}, 287–412, CRM Ser. Math. Phys., Springer, New York, 1999


\bibitem{Fay73} Fay, John D., {\it Theta-functions on Riemann surfaces},
Lect.Notes in Math.,  {\bf 352} Springer (1973)

\bibitem{Fay92} Fay, John D., {\it Kernel functions, analytic torsion, and
moduli spaces}, Memoirs of the AMS {\bf 464} (1992)

\bibitem{GrinOrlov} Grinevich, P., Orlov, A., {\it Flag Spaces in KP Theory and Virasoro Action on det $D_j$ and
Segal-Wilson $\tau$-function}, Preprint CLNS 945/89
Cornell University, 1989, arxiv math-ph/9804019

\bibitem{EKK} Eynard, B., Kokotov, A., Korotkin, D., {\it Genus one contribution to free energy
in Hermitian two-matrix model}, Nuclear Physics {\bf B694} 443-472 (2004)

\bibitem{EKZ} Eskin, A.,  Kontsevich, M., Zorich, A., {\it Sum of Lyapunov exponents of the Hodge bundle with respect to the Teichmuller geodesic flow}, arXiv:1112.5872. 

\bibitem{HarMum}  Harris,  J., Mumford, D., {\it On the Kodaira dimension of the moduli space of curves}, Invent.
Math. {\bf 67}, 23-86 (1982).

\bibitem{HarnadMckay}  Harnad, J.; McKay, J. Modular solutions to equations of generalized Halphen type. R. Soc. Lond. Proc. Ser. A Math. Phys. Eng. Sci. 456 (2000), no. 1994, 261–294

\bibitem{Hitchin} Hitchin, N.
{\it Poncelet polygons and the Painlev\'e equations}. Geometry and analysis (Bombay, 1992), 151–185, Tata Inst. Fund. Res., Bombay, 1995. 

\bibitem{DH} D'Hoker E.,  Phong, D.H.,
{\it Functional determinants on Mandelstam diagrams},
Comm. Math. Phys. {\bf 124} 629--645 (1989)

\bibitem{KalKok} Hillairet, L., Kalvin, V., Kokotov, A., {\it Spectral determinants on Maldelstam diagrams}, 
arXiv:1312.0167

\bibitem{Knizhnik} Knizhnik, V., {\it Multiloop amplitudes in the theory of
quantum strings and complex geometry}, Soviet Phys. Uspekhi {\bf 32}  N11,
945-971 (1989)

\bibitem{IMRN} Kokotov, A.,  Korotkin, D., {\it Isomonodromic tau function of Hurwitz Frobenius manifolds and its applications}, IMRN,  {\bf 2006},  1-34 (2006).

\bibitem{JDG} Kokotov, A., Korotkin, D., {\it Tau functions on spaces of Abelian
differentials and higher genus generalization of Ray-Singer formula},
J. Diff. Geom. {\bf 82}, 35-100 (2009).

\bibitem{Advances} Kokotov, A., Korotkin, D., Zograf, P., {\it Isomonodromic tau
function on the space of admissible covers}, Adv. Math., {\bf 227} no. 1, 586-600 (2011).

\bibitem{KokStr} Kokotov, A., Strachan, I., {\it On the isomonodromic
tau-function for the Hurwitz spaces of branched coverings of genus
zero and one}, Math.Res.Lett.,

\bibitem{Birk} D.Korotkin, {\it Matrix Riemann-Hilbert problems related to branched coverings of $CP^1$},
in ``Operator theory: Advances  and Applications",  103-129,                                                                
ed. by I.Gohberg, A.F. dos Santos  and N.Manojlovic,
Birkhauser, Boston, 2002


\bibitem{Annalen} Korotkin, D., {\it Solution of matrix Riemann-Hilbert problems with
quasi-permutation monodromy matrices},
Math.Ann., {\bf 329} 335-364 (2004)

\bibitem{Crelle} Korotkin, D.,  Shramchenko, V., {\it Inverse monodromy problem for Hurwitz Frobenis manifolds: regular singularities}, J. Reine Agew. Math,
{\bf 2011}, issue 661, 125-187

\bibitem{MRL} Korotkin, D.,  Zograf, P., {\it Tau function and moduli of differentials}, 
Math. Res. Lett. {\bf 18}, no.3, 447-458 (2011).

\bibitem{KonZor}
Kontsevich, M., Zorich,A., {\it Connected components of the moduli spaces
of holomorphic differentials with prescribed
singularities}, Invent. Math. {\bf 153}  631-678 (2003)



\bibitem{Krich} Krichever, I., {\it Integration of non-linear equations by the methods of algebraic geometry, 
Math.USSR Uspekhi}, {\bf 32}. No.6, 183-206 (1977)


\bibitem{Krich1} Krichever, I.M.,  {\it The $\tau$-function of the universal Whitham hierarchy, matrix models and topological field theories.} Comm. Pure Appl. Math. 47 (1994), no. 4, 437-475.

\bibitem{Quillen}
Quillen, D., {\it Determinants of Cauchy-Riemann operators on Riemann surfaces}, Funkts.Anal.Prilozh., {\bf 19} 37-41 (1985)

\bibitem{Rauch}
Rauch, H.E. {\it Weierstrass points, branch points, and moduli of Riemann
surfaces}, Comm. Pure Appl. Math. {\bf 12} 543-560  (1959)

\bibitem{Tata1} Mumford, D. {\it Tata lectures on Theta I},
  Birkh\"auser, Boston (1984)

\bibitem{VerVer} Verlinde, E., Verlinde, H., {\it Chiral bosonization, determinants and the string partition function},
Nucl.Phys. {\bf B288} 357-396 (1987)

\bibitem{Takhtajan} Takhtajan, Leon A. Modular forms as tau-functions for certain integrable reductions of the Yang-Mills equations. Integrable systems (Luminy, 1991), 115-129, Progr. Math., 115, Birkh\"auser Boston, Boston, MA, 1993.

\bibitem{Zorich} Kontsevitch, M., Zorich, A., {\it Lyapunov exponents and
Hodge theory}, arXiv hep-th/9701164

\end{thebibliography}
\end{document}